\documentclass[fleqn,usenatbib]{mnras}

\usepackage{newtxtext,newtxmath}

\usepackage[T1]{fontenc}

\DeclareRobustCommand{\VAN}[3]{#2}
\let\VANthebibliography\thebibliography
\def\thebibliography{\DeclareRobustCommand{\VAN}[3]{##3}\VANthebibliography}

\usepackage{graphicx}	%
\usepackage{amsmath}	%
\usepackage{bm}

\newcommand\varcal[1]{\text{\usefont{OMS}{cmsy}{m}{n}#1}} %

\title[Kernel-marginalised Gaussian processes]{Kernel-, mean- and noise-marginalised Gaussian processes for exoplanet transits and $H_0$ inference}

\author[N. Kroupa et al.]{
Namu Kroupa,$^{1,2,3}$\thanks{E-mail: nk544@cam.ac.uk}
David Yallup,$^{1,2}$
Will Handley$^{1,2}$
and Michael Hobson$^{1}$
\\
$^{1}$Astrophysics Group, Cavendish Laboratory, J. J. Thomson Avenue, Cambridge CB3 0HE, United Kingdom\\
$^{2}$Kavli Institute for Cosmology, Madingley Road, Cambridge CB3 0HA, United Kingdom\\
$^{3}$Engineering Laboratory, University of Cambridge, Cambridge CB2 1PZ, United Kingdom
}

\date{Accepted XXX. Received YYY; in original form ZZZ}

\pubyear{2023}

\begin{document}
\label{firstpage}
\pagerange{\pageref{firstpage}--\pageref{lastpage}}
\maketitle

\begin{abstract}
	Using a fully Bayesian approach, Gaussian Process regression is extended to include marginalisation over 
	the kernel choice and kernel hyperparameters. 
	In addition, Bayesian model comparison via the evidence 
	enables 
	direct kernel comparison. 
	The calculation of the joint posterior was implemented with a transdimensional sampler which simultaneously samples over the discrete kernel choice and their hyperparameters by embedding these in a higher-dimensional space, from which samples are taken using nested sampling. 
	Kernel recovery and mean function inference were explored on synthetic data from exoplanet transit light curve simulations. 
	Subsequently, the method was
	extended to marginalisation over mean functions and noise models and
	applied to the inference of the present-day Hubble parameter, $H_0$, 
	from real measurements of the Hubble parameter as a function of redshift, derived from the 
	cosmologically model-independent 
	cosmic chronometer and 
	$\Lambda$CDM-dependent 
	baryon acoustic oscillation observations.
	The inferred $H_0$ values from the cosmic chronometers, baryon acoustic oscillations and combined datasets are
	$H_0= 66 \pm 6\, \mathrm{km}\,\mathrm{s}^{-1}\,\mathrm{Mpc}^{-1}$, $H_0= 67 \pm 10\, \mathrm{km}\,\mathrm{s}^{-1}\,\mathrm{Mpc}^{-1}$ and $H_0= 69 \pm 6\, \mathrm{km}\,\mathrm{s}^{-1}\,\mathrm{Mpc}^{-1}$, respectively.
	The kernel posterior of the cosmic chronometers dataset prefers a non-stationary linear kernel. Finally, the datasets are shown to be not in tension with $\ln R=12.17\pm 0.02$.

\end{abstract}


\section{Introduction}

Gaussian Processes (GPs) have found widespread application as a method for regression of time-series data in astrophysics, such as the estimation of underlying physical parameters and the modelling of correlated noise, as they offer robust Bayesian non-parametric regression \citep{roberts, aigrain}. 
A GP requires the choice of a kernel, which quantifies the strength of correlation between neighbouring data points.
An existing approach to kernel selection is to write the kernel as a linear combination of several simpler kernels \citep{rasmussen} and let the magnitude of the coefficients determine the relevance of each kernel.
However, in cases where the joint posterior of the coefficients is multimodal or contains curving degeneracies in high dimensions, this lacks interpretability and information can be lost by projecting the posterior down to one- and two-dimensional marginal distributions.
Other approaches allow the kernel to be approximated by kernels which are dense in the set of stationary kernels \citep{wilson} or perform a greedy search over kernel structures \citep{duvenaud2013structure}.
In astrophysical and cosmological datasets, however, the choice of kernels is often governed and restricted by the physical model and interpretability of the hyperparameters in the physical context. 
An example is the selection between physically motivated noise models \citep{scheutwinkel2023bayesian} or the selection between GPs whose form are constrained by solutions to stochastic differential equations \citep{kelly2014flexible}. In this sense, kernel selection can be viewed as a form of Bayesian hypothesis testing, with the Bayes factors determining the kernel to be selected \citep{zhang2023kernel}.
Another approach is deep kernel learning which transforms the kernel inputs with a deep neural network \citep{wilson2016deep}, thus drastically increasing the flexibility of the model.
However, the lack of interpretability of the latent variables of a neural network can be prohibitive. Moreover, sparse and noisy datasets, such as cosmic chronometers and baryon acoustic oscillations, do not meet the demand of such neural networks.
In this paper, kernel selection is approached from a Bayesian perspective by computing the posterior probability of each kernel and sampling from this distribution. This has the advantage of direct interpretability of the kernel posterior probabilities, circumvents the drawbacks outlined above and generalises automatic relevance determination as any linear combination of the investigated kernels can also be included.

A Bayesian approach to model selection requires sampling of the GP hyperparameters to make predictions and calculate marginalised posterior distributions 
\citep{rasmussen, higson, dhawan2021non, simpson}. Most of the current literature approximates this step with a maximum likelihood 
(ML-II) 
estimate 
which is only accurate for a uniform prior and a strongly peaked unimodal likelihood.
Thus, in this paper, we investigate the 
method of marginalising over hyperparameters and the kernel choice and perform model selection of kernels using evidences.
To the knowledge of the authors, this paper is the first to systematically investigate the ability of this method to infer a kernel ground truth and the conditions under which kernel recovery is possible depending on the signal-to-noise ratio and the number of data points.

To sample from the resulting hierarchical model, transdimensional sampling is employed, which samples from the joint posterior of the model choice and the hyperparameter vector of each model. 
The required sampling of the hyperparameters in high dimensions is conventionally done with Markov chain Monte Carlo (MCMC) methods \citep{mackay}. 
A review of transdimensional sampling based on MCMC is given in \citet{sisson2005transdimensional}.
MCMC generates correlated samples and can thus become trapped at a local maximum of the posterior. Hence, these methods can fail to give representative samples and evidence estimates when the posterior distribution is multimodal, contains plateaus or exhibits a phase transition \citep{mackay, hackett}. An alternative method which does not suffer from these shortcomings is nested sampling (NS; \citet{skilling, sivia, ashton, buchner2023nested}), which systematically compresses a set of samples in the parameter space towards the global maximum of the likelihood, continuously generating posterior samples and providing an estimate of the evidence. 
For these reasons, nested sampling is used as the transdimensional sampler implemented in this work \citep{brewer2014inference, hee}.

From the perspective of cosmology and astrophysics, the inference of parameters from data is of primary interest. Thus, kernel inference, corresponding to the inference of the correct noise model, and hyperparameter inference are investigated in exoplanet simulations with correlated noise. These were chosen as exoplanet detection is an active field of research \citep{feroz}, simulation packages exist \citep{foreman-mackey, pytransit} and the shape of the mean function facilitates its separation from noise. In cosmology, one of the most prominent examples for the importance of parameter inference is the discrepancy between early- and late-time measurements of $H_0$, the Hubble tension \citep{valentino, shah, dainotti2022evolution, abdalla, poulin}. Gaussian Process regression (GPR) with a specific kernel and mean function has been used to infer $H_0$ directly from Hubble parameter
measurements \citep{busti2014evidence, o2021elucidating, gomez, bernardo, yu}. 
Here, these methods are extended to kernel, mean function and noise model marginalisation.

The background on GPR and Bayesian model selection 
is summarised in Sections~\ref{sec:gaussian-process-regression} and \ref{sec:bayesian-model-selection}, respectively.
Section~\ref{sec:transdimensional-sampling} details the implementation of the transdimensional sampler.
In Section~\ref{sec:synthetic-data}, the implementation is explored and validated
on synthetic data of exoplanet transits. 
The method is applied to $H_0$ inference 
from real observations of cosmic
chronometers and Baryon acoustic oscillations in Section~\ref{sec:cc-data}.
The paper concludes with Section~\ref{sec:conclusions}. 
Appendices~\ref{appendix:posterior-probability-distribution-from-evidence} and~\ref{appendix:multivariate-gaussian-mixture-distribution} contain derivations of closed-form expressions used to speed up computations, Appendix~\ref{appendix:equivalence-of-linear-kernel-and-linear-mean} shows that a GP with a linear kernel is equivalent to linear regression, Appendix~\ref{appendix:example-of-a-uniform-categorical-prior} illustrates the evidence calculation of a uniform categorical prior, Appendix~\ref{appendix:testing-the-transdimensional-sampler} describes the tests conducted on the implementation of the transdimensional sampler, Appendix~\ref{appendix:prior-ranges} contains the priors used for the kernel hyperparameters and Appendix~\ref{appendix:implementation-of-the-similarity-metric} details the implementation of a novel kernel similarity metric.

\section{Gaussian Process regression}\label{sec:gaussian-process-regression}

A GP is a set of random variables $\{f(\mathbf{x})\mid \mathbf{x}\in \chi\}$, where $\chi$ is an index set\footnote{In Sections~\ref{sec:synthetic-data} and \ref{sec:cc-data}, $\mathbf{x}$ is time, $t$, and redshift, $z$, respectively.}, 
any finite subset of which have a joint Gaussian probability distribution \citep{rasmussen}. It thus describes a probability distribution over functions. 
A GP is specified by a mean function $m(\mathbf{x})$ and a positive definite kernel $k(\mathbf{x}, \mathbf{x'})$, which measures the correlation between $\mathbf{x}$ and $\mathbf{x}'$, such that 
\begin{align}
	m(\mathbf{x})&=\mathbb{E}[f(\mathbf{x})],\\
	k(\mathbf{x} ,\mathbf{x'})&=\mathbb{E}[(f(\mathbf{x})-m(\mathbf{x}))(f(\mathbf{x'})-m(\mathbf{x'}))],
\end{align}
for all $\mathbf{x}$ and $\mathbf{x'}$, where $\mathbb{E}$ denotes the expectation value. 
A kernel is stationary if it depends only on $\mathbf{x}-\mathbf{x}'$, otherwise it is non-stationary.
The combination of a constant mean function and a stationary kernel defines a weakly stationary GP.

The marginalisation property of a GP ensures that any finite subset of the random variables represents the entire GP so that it is not necessary to work with an infinite number of random variables, making GPs computationally tractable.

In GPR, a GP is used as a non-parametric method to perform the regression of a model, $f:\chi\rightarrow\mathbb{R}$, to training inputs, ${X=(\mathbf{x}_1,\dots,\mathbf{x}_{N_\mathrm{data}})}$, and targets, $\mathbf{y}=(y_1,\dots,y_{N_\mathrm{data}})$, to make predictions on test inputs, 
$X^\star=(\mathbf{x}^\star_1,\dots,\mathbf{x}^\star_{n^\star})$.
The targets, $\mathbf{y}$, are assumed to be realisations of the model output, ${\mathbf{f}=(f(\mathbf{x}_1),\dots,f(\mathbf{x}_{N_\mathrm{data}}))}$, with Gaussian noise with covariance matrix, $\mathbf{\Sigma}$,
so that $\mathbf{y}\mid \mathbf{f} \sim \mathcal{N}(\mathbf{f},\mathbf{\Sigma})$,
where $\mathcal{N}$ is the normal distribution and
``$x\sim p(x)$'' denotes that $x$ is sampled from $p(x)$.
Thus, it can be shown 
that the targets, $\mathbf{y}$, and test outputs, $\mathbf{f}^\star=(f(\mathbf{x}_1^\star),\dots,f(\mathbf{x}^\star_{n^\star}))$, are jointly normally distributed,
\begin{equation}\label{eqn:gp-prior}
	\Bigl.\begin{bmatrix}
		\mathbf{y}\\
		\mathbf{f}^\star
	\end{bmatrix}
	\,\Bigr|\, X, X^\star, m, k \sim
	\mathcal{N}\left(
	\begin{bmatrix}\mathbf{m}\\\mathbf{m^\star}\end{bmatrix},
	\begin{bmatrix}
		\mathbf{K}+\mathbf{\Sigma}&\mathbf{K^\star}\\
		\mathbf{K^{\star\top}}&\mathbf{K^{\star\star}}
	\end{bmatrix}
	\right),
\end{equation}
where $\mathbf{m}_i=m(\mathbf{x}_i)$, $\mathbf{m}_i^\star=m(\mathbf{x}_i^\star)$, $\mathbf{K}_{ij}=k(\mathbf{x}_i,\mathbf{x}_j)$, $\mathbf{K}^\star_{ij}=k(\mathbf{x}_i,\mathbf{x}_j^\star)$ and \mbox{$\mathbf{K}^{\star\star}_{ij}=k(\mathbf{x}_i^\star, \mathbf{x}_j^\star)$}.
By conditioning $\text{Equation}$ \ref{eqn:gp-prior} on the targets, the predictive distribution is obtained,
\begin{align}\label{eqn:predictive-distribution}
\begin{split}
	\mathbf{f}^\star\mid\mathbf{y}, X, X^\star, m, k
 \sim\mathcal{N}(&\mathbf{m}^\star+\mathbf{K}^{\star\top}(\mathbf{K}+\mathbf{\Sigma})^{-1}(\mathbf{y}-\mathbf{m}),\\& \ \mathbf{K}^{\star\star}-\mathbf{K}^{\star\top}(\mathbf{K}+\mathbf{\Sigma})^{-1}\mathbf{K}^\star),
 \end{split}
\end{align}
with which predictions at the test inputs, $X^\star$, are computed.

\section{Bayesian model selection}\label{sec:bayesian-model-selection}

It is common in the machine learning literature \citep{rasmussen} to restrict inference to stationary kernels and ${m(\mathbf{x})=0}$.
This choice may not give optimal predictive performance in the low data limit \citep{fortuin2019meta}. In fact, the mean function, $m$, and noise covariance, $\bm{\Sigma}$, are often set by a physical model \citep{aigrain}, which have physical hyperparameters, $\bm{\phi}$ and $\bm{\sigma}$, respectively. 
Hence, this paper focuses on the selection of a kernel from a discrete set of functions, $k_c\in\{k_i\}_{i=1}^{N_\mathrm{kernel}}$, each of which depends on hyperparameters, $\bm{\theta}_c\in\{\bm{\theta}_i\}_{i=1}^{N_\mathrm{kernel}}$, which may have different dimensions,
 with a fixed mean function and noise model.
The approach is applied recursively in Section \ref{sec:cc-data} to include marginalisation over mean functions and noise models.
In the following, the inference of the kernel $k_c$ and the model hyperparameters $\bm{\Theta}_c=(\bm{\theta}_c,\bm{\phi},\bm{\sigma})$ is described in terms of a hierarchical model.

\begin{table*}
	\centering
	\caption{Kernel families commonly used for Gaussian Processes \citep{rasmussen}. $x$ and $x'$ are real numbers and \mbox{$\delta=|x-x'|$}.}
	\label{table:kernels}
	\begin{tabular}{cccl} %
		\hline
		\textbf{Kernel name} & \textbf{Abbreviation} & \textbf{Hyperparameters} & \textbf{Definition}\\
		\hline
		Exponential & E & $A_{\mathrm{E}}$, $\ell_{\mathrm{E}}$&$k(x,x')=A_{\mathrm{E}}^2\exp\left(-\frac{\delta}{\ell_{\mathrm{E}}}\right)$\\
		Matérn-3/2 & M32 & $A_\mathrm{M32}$, $\ell_{\mathrm{M32}}$ & $k(x,x')=A_{\mathrm{M32}}^2\left(1+\frac{\sqrt{3}\delta}{\ell_{\mathrm{M32}}}\right)\exp\left(-\frac{\sqrt{3}\delta}{\ell_{\mathrm{M32}}}\right)$\\
		Matérn-5/2 & M52 & $A_\mathrm{M52}$, $\ell_{\mathrm{M52}}$ & $k(x,x')=A_{\mathrm{M52}}^2\left(1+\frac{\sqrt{5}\delta}{\ell_{\mathrm{M52}}}+\frac{5\delta^2}{3\ell^2_{\mathrm{M52}}}\right)\exp\left(-\frac{\sqrt{5}\delta}{\ell_{\mathrm{M52}}}\right)$\\
  Matérn-7/2&M72&$A_\mathrm{M72}$, $\ell_{\mathrm{M72}}$&$k(x,x')=A_{\mathrm{M72}}^2\left(1+\frac{\sqrt{7}\delta}{\ell_{\mathrm{M72}}}+\frac{14\delta^2}{5\ell_{\mathrm{M72}}^2}+\frac{7\sqrt{7}\delta^3}{15\ell_\mathrm{M72}^3}\right)\exp\left(-\frac{\sqrt{7}\delta}{\ell_{\mathrm{M72}}}\right)$\\
  Squared exponential&SE&$A_{\mathrm{SE}}$, $\ell_{\mathrm{SE}}$&$k(x,x')=A_{\mathrm{SE}}^2\exp\left(-\frac{\delta^2}{2\ell_{\mathrm{SE}}^2}\right)$\\
  Rational quadratic&RQ&$A_{\mathrm{RQ}}$, $\ell_{\mathrm{RQ}}$, $\alpha$&$k(x,x')=A_{\mathrm{RQ}}^2\left(1+\frac{\delta^2}{2\ell_{\mathrm{RQ}}^2\alpha}\right)^{-\alpha}$\\
  Exponential sine squared&ESS&$A_{\mathrm{ESS}}$, $\Gamma$, $P_{\mathrm{ESS}}$&$k(x,x')=A_{\mathrm{ESS}}^2\exp\left(-\Gamma\sin^2\left(\frac{\pi\delta}{P_{\mathrm{ESS}}}\right)\right)$\\
  Cosine&Cos&$A_{\mathrm{Cos}}$, $P_{\mathrm{Cos}}$&$k(x,x')=A_{\mathrm{Cos}}^2\cos\left(\frac{2\pi\delta}{P_{\mathrm{Cos}}}\right)$\\
  Linear&L&$A_\text{L,1}$, $A_\text{L,2}$&$k(x,x')=A_\text{L,1}^2+A_\text{L,2}^2xx'$\\
		\hline
	\end{tabular}
\end{table*}

\subsection{Hyperparameter selection}\label{sec:hyperparameter-selection}

For a given choice of the kernel, $k_c$, the hyperparameters are subject to a posterior, which is the probability distribution conditioned on the observation of training data, $(\mathbf{y},X)$. This is given by Bayes theorem as
\begin{equation}\label{eqn:hyperparameter-posterior-distribution}
	\varcal{P}_{k_c}(\bm{\Theta}_c)=\frac{\varcal{L}_{k_c}(\bm{\Theta}_c)\pi_{k_c}(\bm{\Theta}_c)}{\varcal{Z}_{k_c}},
\end{equation}
where $\pi_{k_c}(\bm{\Theta}_c)$ is the hyperparameter prior, which encodes any information on $\bm{\Theta}_c$ prior to the observation of data. The single-kernel likelihood, $\varcal{L}_{k_c}$, is given 
by 
\begin{equation}\label{eqn:gp-likelihood}
\begin{split}
	\varcal{L}_{k_c}(\bm{\Theta}_c)=&\frac{1}{\sqrt{|2\pi(\mathbf{K}+\bm{\Sigma})|}}\\& \times\exp\left(-\frac12(\mathbf{y}-\mathbf{m})^\top(\mathbf{K}+\bm{\Sigma})^{-1}(\mathbf{y}-\mathbf{m})\right),
 \end{split}
\end{equation}
and the kernel evidence, $\varcal{Z}_{k_c}$, 
in $\text{Equation}$ \ref{eqn:hyperparameter-posterior-distribution} is
\begin{equation}\label{eqn:hyperparameter-evidence}
	\varcal{Z}_{k_c}=\int \varcal{L}_{k_c}(\bm{\Theta}_c)\pi_{k_c}(\mathbf{\Theta}_c)\mathrm{d}\mathbf{\Theta}_c.
\end{equation}
Unless stated otherwise, $\pi_{k_c}(\mathbf{\Theta}_c)$ is assumed to be a weakly informative uniform prior, $\varcal{U}(a,b)$, where $[a,b]$ is an interval.

\subsection{Kernel selection}

The kernel posterior for the choice from $N_\mathrm{kernel}$ kernels 
is given by Bayes theorem as
\begin{equation}\label{eqn:kernel-posterior-distribution}
	p_{k_c}=\frac{\varcal{Z}_{k_c}\Pi_{k_c}}{\varcal{Z}},
\end{equation}
where $\Pi_{k_c}$ is the prior over the kernel choice and the evidence, $\varcal{Z}$, is 
\begin{equation}\label{eqn:kernel-evidence}
	\varcal{Z}=\sum_{i=1}^{N_\mathrm{kernel}} \varcal{Z}_{k_i}\Pi_{k_c}.
\end{equation}
The prior $\Pi_{k_c}$ is taken to be uniform, $\Pi_{k_c}=\frac{1}{N_\mathrm{kernel}}$.
An expression for the propagated uncertainty in $p_{k_c}$ is given in Appendix~\ref{appendix:posterior-probability-distribution-from-evidence}.

\subsection{Inference}

Thus, given a dataset, $X$ and $\mathbf{y}$, and priors $\{\pi_{k_c}(\bm{\Theta}_c)\}$, the kernel and hyperparameters can be inferred from Equations \ref{eqn:hyperparameter-posterior-distribution} and \ref{eqn:kernel-posterior-distribution}. The posterior of any quantity, $Q$, which depends on the kernel and its hyperparameters is obtained by marginalising over them \citep{simpson}:
\begin{equation}\label{eqn:inference-of-quantity}
	p(Q\mid\mathbf{y}, X)=\sum_{i=1}^{N_\mathrm{kernel}}\int p(Q\mid k_i, \bm{\Theta}_i)p_{k_i}\varcal{P}_{k_i}(\bm{\Theta}_i)\mathrm{d}\bm{\Theta}_i.
\end{equation}
Since it is impossible to calculate this in general, one instead samples so that
\begin{equation}\label{eqn:quantity-sample-from-distribution}
	p(Q\mid\mathbf{y}, X)\approx \frac{1}{M}\sum_{j=1}^Mp(Q\mid k^{(j)}, \bm{\Theta}^{(j)}),
\end{equation}
where $\{(k^{(j)}, \bm{\Theta}^{(j)})\}$ are $M$ samples from the joint posterior, $p_{k_i}\varcal{P}_{k_i}(\bm{\Theta})$. If the distributions $p(Q\mid k^{(j)},\bm{\Theta}^{(j)})$ are Gaussian, closed-form expressions for the mean and covariance of $p(Q\mid\mathbf{y},X)$ can be derived (Appendix~\ref{appendix:multivariate-gaussian-mixture-distribution}).

\subsection{Common kernel choices}

Table~\ref{table:kernels} lists the set of kernels used throughout this paper.
The exponential (E), Mat\'ern-3/2 (M32), Mat\'ern-5/2 (M52), Mat\'ern-7/2 (M72) and squared exponential (SE) kernels are specific instances of the Mat\'ern kernel family. Functions sampled from their corresponding GPs are increasingly smoother in the order listed, with the E kernel being zero times and the SE kernel infinitely many times differentiable in the mean square sense \citep{rasmussen}. Each of them has an amplitude and length scale hyperparameter, which is interpretable as a decorrelation length on the input space.
The rational quadratic (RQ) kernel corresponds to a mixture of SE kernels with different length scales, where the hyperparameter $\alpha$ controls the contributions of the length scales.
Functions sampled from the GPs of the exponential sine squared (ESS) and cosine (Cos) kernels are periodic with period $P_\mathrm{ESS}$ and $P_\mathrm{Cos}$, respectively.
Finally, using the non-stationary linear (L) kernel is equivalent to performing linear regression with Gaussian priors on the offset and slope with zero mean and variances $A^2_\mathrm{L,1}$ and $A^2_\mathrm{L,2}$, respectively, which in turn is equivalent to a GP with a linear mean function (Appendix~\ref{appendix:equivalence-of-linear-kernel-and-linear-mean}).

\subsection{Kullback-Leibler divergence and Bayesian model dimensionality}

The Kullback-Leibler (KL) divergence of the hierarchical model above is defined as an average over the joint posterior of the kernel choice and the kernel hyperparameters,
\begin{equation}\label{eqn:KL-div-1}
	\varcal{D}_\mathrm{KL}=\underset{(k_c, \bm{\Theta}_c)\sim p_{k_c}\varcal{P}_{k_c}(\bm{\Theta}_c)}{\mathbb{E}}\left[\ln\frac{p_{k_c}\varcal{P}_{k_c}(\bm{\Theta}_c)}{\Pi_{k_c}\pi_{k_c}(\bm{\Theta}_c)}\right],
\end{equation}
which decomposes into contributions from a KL divergence arising purely from the constraint on the kernel choice and an averaged KL divergence of the individual kernel hyperparameters,
\begin{equation}\label{eqn:KL-div-2}
\begin{split}
	\varcal{D}_\mathrm{KL}=\ &\underset{k_c\sim p_{k_c}}{{\mathbb{E}}}\left[\ln \frac{p_{k_c}}{\Pi_{k_c}}\right]\\
	&+\underset{k_c\sim p_{k_c}}{\mathbb{E}}\left[\underset{\bm{\Theta}_c\sim\varcal{P}_{k_c}(\bm{\Theta}_c)}{\mathbb{E}}\left[\ln\frac{\varcal{P}_{k_c}(\bm{\Theta}_c)}{\pi_{k_c}(\bm{\Theta}_c)}\right]\right].
\end{split}
\end{equation}

The Bayesian model dimensionality $d_B$ (BMD; \citet{handley-2}) is defined as 
\begin{equation}\label{eqn:bmd-1}
	\frac{d_B}{2}=\underset{(k_c,\bm{\Theta}_c)\sim p_{k_c}\varcal{P}_{k_c}(\bm{\Theta}_c)}{\mathrm{Var}}\left[\ln\frac{p_{k_c}\varcal{P}_{k_c}(\bm{\Theta}_c)}{\Pi_{k_c}\pi_{k_c}(\bm{\Theta}_c)}\right],
\end{equation}
where $\mathrm{Var}[\cdot]$ denotes the variance, and is the sum of BMDs arising from the constraint of the kernel posterior and the hyperparameter posteriors with an additional term increasing (decreasing) the BMD if the kernel posterior compression is correlated (anti-correlated) with the compression of the corresponding hyperparameter posterior,
\begin{equation}\label{eqn:bmd-2}
\begin{split}
	\frac{d_B}{2}=&\underset{k_c\sim p_{k_c}}{\mathrm{Var}}\left[\ln \frac{p_{k_c}}{\Pi_{k_c}}\right]\\
	+&\underset{(k_c,\bm{\Theta}_c)\sim p_{k_c}\varcal{P}_{k_c}(\bm{\Theta}_c)}{\mathrm{Var}}\left[\ln\frac{\varcal{P}_{k_c}(\bm{\Theta}_c)}{\pi_{k_c}(\bm{\Theta}_c)}\right]\\
	+&\,2\underset{(k_c,\bm{\Theta}_c)\sim p_{k_c}\varcal{P}_{k_c}(\bm{\Theta}_c)}{\mathrm{Cov}}\left[\ln \frac{p_{k_c}}{\Pi_{k_c}},\,\ln\frac{\varcal{P}_{k_c}(\bm{\Theta}_c)}{\pi_{k_c}(\bm{\Theta}_c)}\right],
\end{split}
\end{equation}
where $\mathrm{Cov}[\cdot,\cdot]$ denotes the covariance.

The above expressions for $\varcal{D}_\mathrm{KL}$ and $d_B$ apply to any hierarchical model and will be applied recursively in Section~\ref{sec:cc-data} to marginalise over the mean function and noise model.

\section{Transdimensional sampling}\label{sec:transdimensional-sampling}

A valid approach to compute the posterior probabilities $p_{k_c}$ 
is to fit each GP model sequentially to the data and to compute the evidence $\varcal{Z}_{k_c}$ for each.
A subsequent calculation of $Q$ requires two-step sampling, firstly of a model choice from $p_{k_c}$ and secondly of the hyperparameters from $\varcal{P}_{k_c}(\bm{\Theta}_c)$ for the given choice.

A computationally more efficient method is to use transdimensional sampling \citep{brewer2014inference, hee}, in which the hyperparameters for each model are embedded in a higher dimensional space such that the dimension of the models match and the model choice is promoted to a categorical hyperparameter, $c$.
In the following, the method of \citet{hee} is adopted. There are two parts to the transdimensional sampling algorithm. Firstly, the appropriate hyperparameter vector must be prepared. Secondly, NS 
is used as a subroutine to sample from this hyperparameter vector,
for which the \textsc{PolyChord} implementation is used \citep{handley}.
Specifically, for the investigated datasets, the models corresponding to different kernels share the same mean function, $m$, and noise model, $\bm{\Sigma}$. The combined hyperparameter vector then takes the form 
\begin{equation}\label{eqn:transdimensional-hyperparameter-vector}
\bm{\Phi}=(c,\bm{\phi},\bm{\sigma},\bm{\theta}_1,\bm{\theta}_2,\dots,\bm{\theta}_{N_\mathrm{kernel}}), 
\end{equation}
where the categorical hyperparameter, $c$, takes values in $\{1,\dots,N_\mathrm{kernel}\}$ and $\bm{\phi}$, $\bm{\sigma}$, and $\{\bm{\theta}_i\}$ are the hyperparameters of the mean function, noise and kernels, respectively. The likelihood, $\varcal{L}$, reduces to the single-kernel likelihood, $\varcal{L}_{k_c}$, ($\text{Equation}$ \ref{eqn:gp-likelihood}) for the kernel $k_c$ selected by $c$, $\varcal{L}(\bm{\Phi})=\varcal{L}_{k_c}(\bm{\phi}, \bm{\sigma}, \bm{\theta}_c)$.
The prior on $\bm{\Phi}$ is the product of uniform priors on each of the entries of $\bm{\Phi}$, unless stated otherwise.
The priors are specified in \textsc{PolyChord} as inverse transforms on the unit hypercube \citep{handley}. 
For the discrete hyperparameter $c$, the function $u\mapsto \lceil N_\mathrm{kernel} u \rceil$ achieves the required uniform categorical distribution, where $u\in [0, 1]$ and $\lceil \cdot \rceil$ is the ceiling function\footnote{In the computational implementation, $c$ takes values in $\{0, \dots, N_\mathrm{kernel}-1\}$ so that the inverse transform must be modified to $u\mapsto  \lfloor N_\mathrm{kernel} u \rfloor$, where $\lfloor\cdot\rfloor$ is the floor function.}. 
The effect is that the hypercube coordinate of $c$ is partitioned into $N_\mathrm{kernel}$ intervals of equal length, each interval corresponding to a value of $c$. This is illustrated in Appendix~\ref{appendix:example-of-a-uniform-categorical-prior}. As a result, the likelihood in unit hypercube coordinates is piecewise constant on each interval, assuming that all other hyperparameters are held fixed, which necessitates a modified NS algorithm or correction by sample post-processing in the \textsc{anesthetic} library \citep{fowlie2021nested}.

At a single NS iteration, the value of $c$ selects the entries $\bm{\Theta}_c$ in $\bm{\Phi}$ which the likelihood $\varcal{L}(\bm{\Phi})$ depends on. The likelihood is degenerate in the remaining entries, $\bm{\theta}_{\bar{c}}=(\bm{\theta}_1,\dots,\bm{\theta}_{c-1},\bm{\theta}_{c+1},\dots,\bm{\theta}_{N_\mathrm{kernel}})$, so that NS does not compress the live points in the subspace of $\bm{\theta}_{\bar{c}}$.
As a consequence, the distribution of $\bm{\theta}_{\bar{c}}$ remains equal to the prior, which in our case is uniform.
Hence, a single NS
run in the space of $\bm{\Phi}$ 
produces samples from the joint posterior over the model choice and the hyperparameters, $\varcal{P}(\bm{\Phi})=p_{k_c}\varcal{P}_{k_c}(\bm{\Theta}_c)\pi_{k_c}(\bm{\theta}_{\bar{c}})$, and the evidence $\varcal{Z}$.
The marginal distribution of $c$ thus corresponds directly to the kernel posterior. To obtain the hyperparameter posterior of a kernel, $\varcal{P}(\bm{\Phi})$ is conditioned on a value of $c$ and marginalised over $\bm{\theta}_{\bar{c}}$.
Computationally, this corresponds to discarding all samples which have a different value of $c$ and computing the histogram of only the entries in $\bm{\Phi}$ corresponding to $\bm{\Theta}_c$.

An alternative approach to compute $p_{k_c}$ is to split the samples according to the value of $c$. Each set of samples
constitutes a separate NS run with a potentially variable number of live points, which must be reweighted \citep{higson2019dynamic} to calculate the kernel evidence $\varcal{Z}_{k_c}$ from which the kernel posterior is obtained.
We note that the hyperparameters $c$ and $\bm{\theta}_{\bar{c}}$ do not affect the evidence calculation as the kernel likelihood $\varcal{L}_{k_c}$ is independent of them.

There are in principle two other ways of combining the individual hyperparameter vectors for transdimensional sampling. 
The first is in keeping with taking the disjoint union of the model parameter spaces, such as in reversible-jump MCMC \citep{green1995reversible}.
That is, we could define the combined vector as $(c,\bm{\phi},\bm{\sigma},\bm{\theta}')$, where $\bm{\theta}'$ is a vector with the maximum dimensionality of the vectors $\{\bm{\theta}_i\}_{i=1}^{N_\mathrm{kernel}}$. This has the advantage of a reduced dimensionality of the hyperparameter space although one cannot apply different priors for the hyperparameters of different kernels when using NS.
Secondly, a product space approach could be taken \citep{carlin1995bayesian}.
That is, the combined vector could be defined as $(c,\bm{\Theta}_1,\dots,\bm{\Theta}_{N_\mathrm{kernel}})$, 
with which each hyperparameter can be assigned a different prior, however the dimensionality is increased by having separate vector entries for the mean functions and noise models of each GP model.
The approach that was taken in this paper corresponds to an adaptive approach between the two \citep{godsill2001relationship, hee}. To wit, we take the first approach for the shared hyperparameters, $\bm{\phi}$ and $\bm{\sigma}$, and the second one for the hyperparameters not shared between the models, $\{\bm{\theta}\}_{i=1}^{N_\mathrm{kernel}}$, thus resulting in the combined vector of Equation~\ref{eqn:transdimensional-hyperparameter-vector}.
This method 
is advantageous for the applications in this paper as each hyperparameter can be assigned a different prior and the number of dimensions of the sampling space used for the mean function and the noise hyperparameters does not scale with the number of kernels. On the downside, the dimension of the sampling space scales with total number of kernel hyperparameters.

We note that regardless of the implementation of transdimensional sampling, Equations \ref{eqn:KL-div-2} and \ref{eqn:bmd-2} remain valid ways of calculating the KL divergence and BMD, respectively, because the posterior and prior, $\pi_{k_c}(\bm{\theta}_{\bar{c}})$, cancel in the posterior to prior ratio. However, it is computationally more efficient to calculate these directly from the definitions, Equations \ref{eqn:KL-div-1} and \ref{eqn:bmd-1}, as one has direct access to samples from the joint posterior by means of $\varcal{P}(\bm{\Phi})$.
As a corollary, the effective (Bayesian) dimensionality of the space of $\bm{\Phi}$ is independent of the transdimensional sampling method.

The transdimensional sampler (TS) was implemented 
using \textsc{jax} \citep{frostig}, leveraging the speedup of just-in-time compilation when evaluated repeatedly. The implementation of $\text{Equation}$~\ref{eqn:gp-likelihood} in \textsc{tinygp} \citep{tinygp} was used due to its compatibility with \textsc{jax}.
The \textsc{anesthetic} library \citep{handley-anesthetic} was used for NS sample post-processing and evidence, KL divergence and BMD calculations.
Since the live points are initially equally partitioned between the subspaces of fixed $c$, the default number of live points in \textsc{PolyChord} was increased by a factor of $N_\mathrm{kernel}$.

The time complexity of a single evaluation of $\varcal{L}(\bm{\Phi})$
is $\varcal{O}(N_\mathrm{data}^3)$ \citep{rasmussen} so that a full \textsc{PolyChord} run has a time complexity $\varcal{O}(N_\mathrm{dims}^3N_\mathrm{data}^3)$ \citep{handley}, where $N_\mathrm{dims}$ is the number of dimensions of $\bm{\Phi}$. 
The computations were 
parallelised with \textsc{mpi4py} \citep{mpi4py}. Tests for the TS are described in Appendix~\ref{appendix:testing-the-transdimensional-sampler}.

\section{Synthetic exoplanet transit data}\label{sec:synthetic-data}

In this Section, the method is validated on synthetic datasets simulated from a known mean function and kernel. Firstly, we test under which conditions the true kernel can be inferred given the mean function. Secondly, we fit for both the mean function and the kernels in Table~\ref{table:kernels} and investigate the accuracy of the mean function hyperparameter inference, which is of greater interest from a physics perspective.

The mean function is a light curve simulated from an exoplanet transit \citep{winn}, which was chosen as the shape thereof facilitates its separation from the noise of an M32 kernel.
Moreover, inference of the transit light curve parameters is of greater interest as it requires that correlated noise is accounted for explicitly \citep{aigrain}, for which GPs have been employed \citep{gibson}.
Exoplanet transit light curves were simulated using the \textsc{jaxoplanet} library \citep{jaxoplanet, aigrain}.
In the following, the creation of the datasets for kernel inference and mean function hyperparameter inference is described.

\subsection{Method}\label{sec:synthetic-data-method}

For kernel inference, a quadratically limb darkened light curve on the time interval $[-0.2\,\mathrm{days},0.2\,\mathrm{days}]$ was used as the mean function for a GP with correlated noise from an M32 kernel and a white noise term, $\bm{\Sigma}=\sigma^2\mathbf{I}$, where $\sigma$ is a hyperparameter and $\mathbf{I}$ is the identity matrix, so that the synthetic data is sampled from a Gaussian $\mathcal{N}(\mathbf{m},\mathbf{K}+\sigma^2\mathbf{I})$.
The simulation hyperparameters are shown in Table~\ref{table:exoplanet-simulation-parameters}. 
Next, we define the signal-to-noise ratio as the ratio of the kernel amplitude and the white noise, $\mathrm{SNR}=A_\mathrm{M32}/\sigma$.
By varying the number of data points $N_\mathrm{data}\in\{75,\dots,750\}$ in steps of $75$ and the signal-to-noise ratio $\log_{10}(\mathrm{SNR})\in\{0,\dots,1\}$ in steps of $0.05$ by keeping $A_\mathrm{M32}$ fixed and varying $\sigma$, a series of datasets was created. For each of these, GPR was performed for every kernel in Table~\ref{table:kernels} with the mean function set to the true mean function.
When the mean function was fitted to the data, the hyperparameter priors of the mean function were chosen to be sufficiently narrow to achieve fast convergence to the posterior bulk during a NS run, yet wide enough to prevent significant truncation thereof.
The kernel hyperparameter priors are described in Appendix~\ref{appendix:prior-ranges}. The prior on $\sigma$ is the same as on the amplitude $A$ of any kernel. Finally, for each dataset the kernel posterior $p_k$ and the hyperparameter posteriors are calculated using the TS (Section~\ref{sec:transdimensional-sampling}).

\begin{figure}
	\centering
	\includegraphics{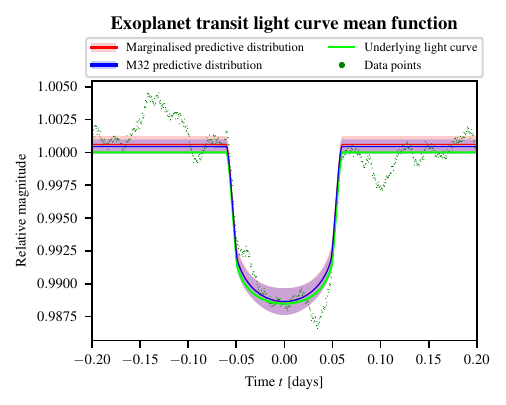}
	\vspace{-2em}
	\caption{Example synthetic dataset for 
		$N_\mathrm{data}=750$
		and $\log_{10}(\mathrm{SNR})=1$. The green curve is the true mean function calculated from an exoplanet transit light curve simulation. Adding noise from an M32 kernel, the data points are obtained. 
		The red and blue curves are the mean function predictive distributions, marginalised over the kernel posterior and conditioned on the M32 kernel, respectively.
		The shaded regions are one-sigma
		error bands. 
	}
\label{fig:mean-function-example}
\end{figure}

For mean function hyperparameter inference, the datasets with $N_\mathrm{data}=75$ and $\log_{10}(\mathrm{SNR})\in\{0,\dots,2\}$ in steps of $0.5$ are created. GPR is performed with $P_\mathrm{orb}$ fixed to the true value because the datasets only contain a single occultation so that $P_\mathrm{orb}$ cannot be inferred. Performing GPR with the M32, M52, E and SE kernels showed that the $q_2$ and $b$ hyperparameters remain unconstrained. Hence, in addition to the datasets described above, GPR is re-performed with $q_2$ and $b$ fixed to their true values.
An example dataset and the resulting mean function predictive distributions are shown in Figure~\ref{fig:mean-function-example}, calculated for $q_2$ and $b$ fitted freely. 

\begin{table}
	\centering
	\caption{True values of the hyperparameters used for the creation of the synthetic datasets and the priors used for 
		hyperparameter inference. The priors marked with (*) are given in Appendix~\ref{appendix:prior-ranges}.}
	\label{table:exoplanet-simulation-parameters}
	\begin{tabular}{ccc}
		\hline
		\textbf{Hyperparameter}&\textbf{True value}&\textbf{Prior}\\
		\hline 
		$q_1$&0.25&$\varcal{U}(0,1)$\\
		$q_2$&0.3&$\varcal{U}(0,1)$\\
		$P_\mathrm{orb}\ [\text{day}]$&1&$-$\\
		$T\ [\text{day}]$&0.12&$\varcal{U}(0.10,0.14)$\\
		$t_0\ [\text{day}]$&0&$\varcal{U}(-0.1,0.1)$\\
		$b$&0.1&$\varcal{U}(0,0.2)$\\
		$r$&0.1&$\varcal{U}(0.05,0.15)$\\
		$f_0$&1&$\varcal{U}(0.9,1.1)$\\
		$A_\mathrm{M32}$&0.002&(*)\\
		$\ell_\mathrm{M32}\ [\text{day}]$&0.02&(*)\\
		\hline
	\end{tabular}
\end{table}

\subsection{Kernel inference results}\label{sec:kernel-inference-results}

\begin{figure*}
		\centering
		\includegraphics{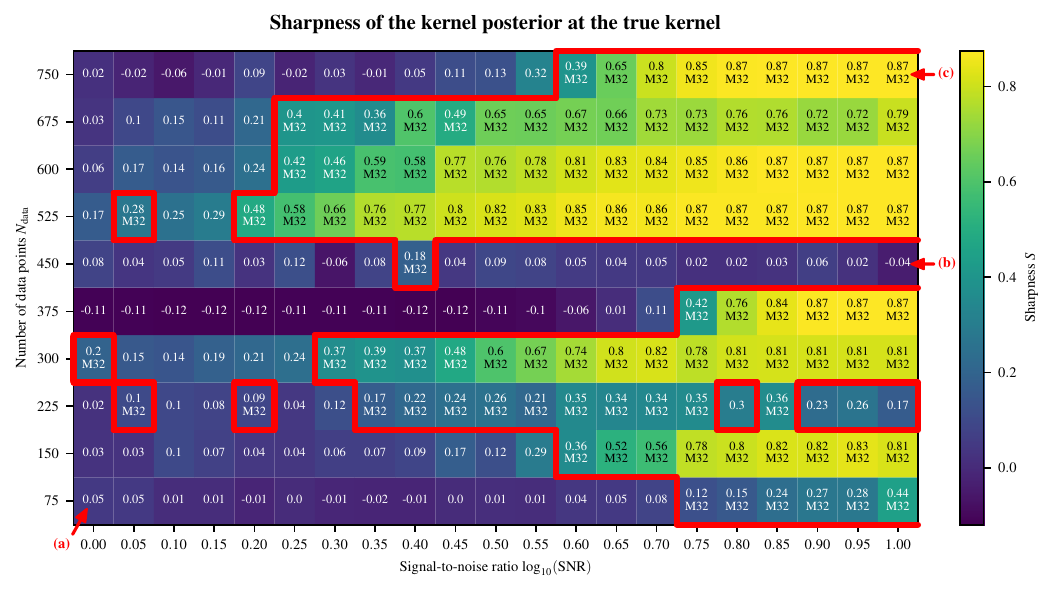}
		\vspace{-3em}
		\caption{Plot showing that the inference of the true kernel is correlated with the noise level and the number of data points. A measure of the sharpness of the kernel posterior at the M32 kernel, $S=p_\text{M32}-\langle p_k \rangle$, is plotted against the signal-to-noise ratio, $\log_{10}(\text{SNR})$, and the number of data points, $N_\text{data}$, of synthetic datasets created from an exoplanet simulation (Section \ref{sec:synthetic-data-method}) and correlated noise from an M32 kernel. Each coloured
			square 
			corresponds to a dataset and the inscribed upper value is the plotted value of 
			$S$, and the lower text is the maximum a posteriori (MAP) kernel. In the region within the red border, the true M32 kernel maximises the posterior. For $N_\text{data}=225$, $450$ and $675$, the M52 kernel has a similar or larger posterior probability than the M32 kernel even for $\log_{10}(\text{SNR})\approx 1$. The complete kernel posteriors for the datasets marked (a), (b) and (c) are shown in $\text{Figure}$ \ref{fig:kernel-posterior-examples}. 
		}
		\label{fig:contour-plot}
\end{figure*}

\begin{figure}
	\centering
	\includegraphics{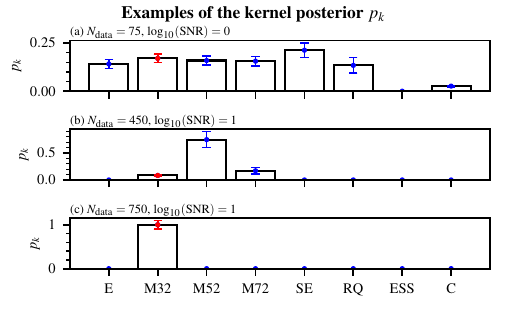}
	\vspace{-3em}
	\caption{Example kernel posteriors showing the inference of the true M32 kernel for the datasets marked (a), (b) and (c) in $\text{Figure}$ \ref{fig:contour-plot}. In each plot, the red data point indicates the true kernel. \textbf{(a)} This is in the low data point, high noise region. The kernel posterior shows that multiple kernels are equally probable within one sigma.
		\textbf{(b)} This is in the medium data point, low noise region. The posterior favours the M52 and M72 kernels compared to the M32 kernel. \textbf{(c)} This is in the high data point, low noise region. The M32 kernel is correctly inferred.}
	\label{fig:kernel-posterior-examples}
\end{figure}

We define a measure of the sharpness of the kernel posterior as ${S=p_\mathrm{M32}-\langle p_k\rangle}$, where $\langle p_k\rangle=1/N_\mathrm{kernel}$ is the average kernel posterior probability. This is large when the kernel posterior is unimodal and sharply peaked at the true kernel and close to zero when the kernel posterior is flat. An alternative such measure is the entropy, $-\Sigma_kp_k\ln p_k$, 
which attains a minimum value of zero when $p_k=1$ for a single kernel and a maximum value of $\ln N_\mathrm{kernel}$ when the kernel posterior is flat.
However, we intend to isolate the behaviour of the posterior at the true kernel so that $S$ is used. The dependence of $S$ on $N_\mathrm{data}$ and $\log_{10}(\mathrm{SNR})$ is shown in Figure~\ref{fig:contour-plot}, in which the region in which the M32 kernel is the maximum a posteriori (MAP) kernel is demarcated with a red border.
Overall, the M32 kernel is recovered at low noise ($\log_{10}(\text{SNR})\gtrsim 0.5$) but not for high noise ($\log_{10}(\text{SNR})\lesssim 0.5$) and this is correlated with the sharpness $S$ of the kernel posterior.
Thus, $S$
is a measure for recovering the true kernel in the sense that a sharply peaked kernel posterior indicates that the true kernel is recovered and, conversely, a flat posterior indicates the opposite.
For closer investigation, the kernel posteriors of the datasets marked (a), (b) and (c) in Figure~\ref{fig:contour-plot} are shown in Figure~\ref{fig:kernel-posterior-examples}. 

Furthermore, it is observed that, in the low SNR
region (large $\sigma$), the M32 kernel cannot be recovered despite $N_\text{data}$ being increased. 
Instead, the kernel posterior is flat with an entropy near $\ln N_\mathrm{kernel}$, indicating that the dataset does not contain sufficient information to constrain the kernel prior.
The reason for the flatness is that the covariance matrix, $\mathbf{K}+\sigma^2\mathbf{I}$, of the normal distribution the data is sampled from receives a significant contribution from the 
diagonal white noise term, $\sigma^2\mathbf{I}$, which decreases the correlation of the samples.
Hence, the true kernel cannot be inferred for high noise even if $N_\text{data}$ is increased. 
Visually, $\log_{10}(\mathrm{SNR})=0$ corresponds to a dataset with almost no discernible pattern in the scatter of the data points, which is reflected in the flat kernel posterior.

\begin{figure}
	\centering
	\includegraphics{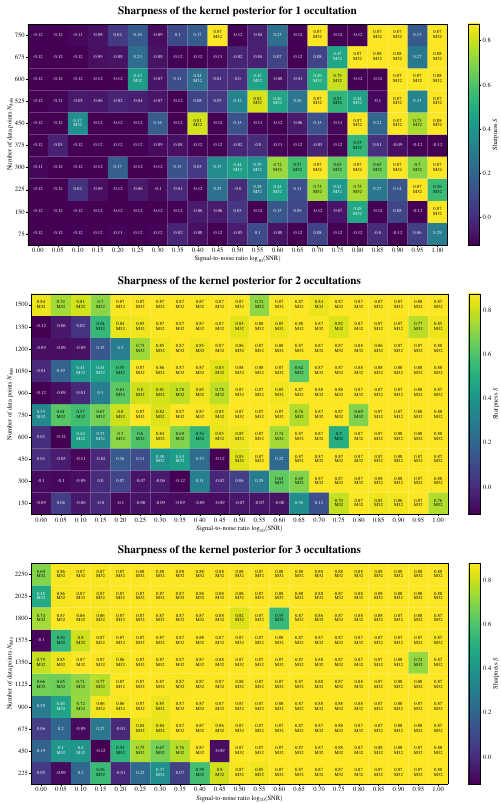}
	\caption{
		Plot of the sharpness, $S$, when multiple occultations are included in the dataset and the mean function and kernels are fit to the data simultaneously.
		As the number of occultations in the dataset is increased, the lower left triangular region in which the kernel cannot be inferred shrinks while the upper right region in which the true kernel can be inferred grows.
		This is consistent with the results for a fixed mean function (Figure~\ref{fig:contour-plot}) and shows that the method is robust for a free mean function when statistically independent realisations of an occultation are present in the dataset.
	}
	\label{fig:multiple-occultations-contour-plots}
\end{figure}

It is expected that the plot in Figure~\ref{fig:contour-plot} roughly separates into two regions. Firstly, a region in which the true kernel cannot be inferred and $S$ is low, corresponding to low SNR and $N_\mathrm{data}$, and secondly a region in which the true kernel is recovered and $S$ is high, corresponding to high SNR and $N_\mathrm{data}$. Moreover, these regions should be separated by a diagonal border.
Figure~\ref{fig:contour-plot} shows multiple deviations from this expectation:
\begin{enumerate}
\item For $N_\mathrm{data}\in\{225, 450, 675\}$, at high SNR, the true kernel is not correctly inferred. Instead, the kernel posterior shows a preference for the M52 and M72 kernels.
\item For $N_\mathrm{data}\in\{375,750\}$, up to ${\log_{10}(\mathrm{SNR})\approx 0.70}$, $S$ remains low. Inspection of the kernel posterior shows that the E kernel is favoured.
\end{enumerate}
These deviations are removed by extending the dataset to include multiple occultations, yielding Figure~\ref{fig:multiple-occultations-contour-plots} which is discussed further below. We therefore conclude that these outliers are statistical fluctuations arising from the finite size of the dataset which do not occur in a real application which typically includes multiple occultations.

In the following, we show that deviations (i) and (ii) persist under the injection and removal of different types of noise which validate this conclusion.
Firstly, we rule out that the deviations are caused by the cadence of the sampling of the light curve. For this, the observation times of the light curve are randomly shifted by adding independent and identically distributed uniform noise on the inputs. The deviations persist and are thus a feature intrinsic to the shape of the light curve.
Secondly, we calculate $S$ in the extreme case of infinite SNR, realised by setting $\sigma=0$. In this limit, the M32 kernel is the MAP kernel for all $N_\mathrm{data}$ except for $N_\mathrm{data}=225$, for which the posterior probability is finite for the M32 and M52 kernels. From this, we validate that the ambiguity between the M32 and M52 kernels is a feature of the shape of the light curve for $N_\mathrm{data}=225$, whereas for all other values of $N_\mathrm{data}$, it is caused by the noise term, $\sigma^2\mathbf{I}$.
Thirdly, by changing the random number generator seed to generate a different dataset and replotting $S$ we conclude that if $S$ is averaged over different realisations of the dataset, the deviations disappear. As the average over several realisations of the same light curve incorporates statistically independent examples of occultations, this is consistent with the finite size of the dataset causing the deviations.

\begin{figure}
	\centering
	\includegraphics{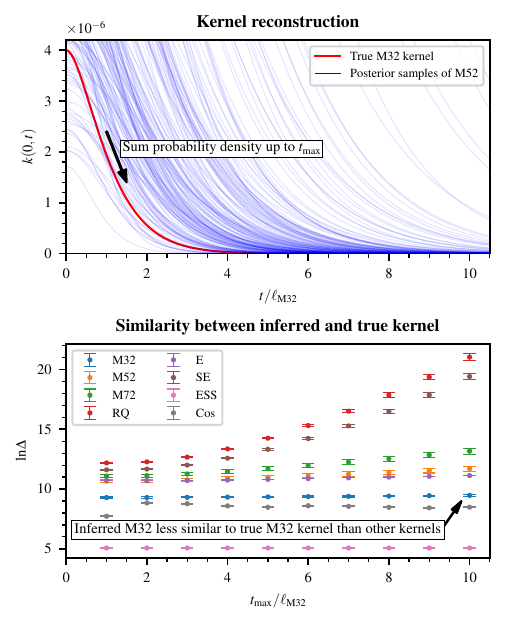}
	\vspace{-3em}
	\caption{
		\textbf{Top:} Plot of the true kernel, which is an M32 kernel with ${A_\mathrm{M32}=0.002}$ and $\ell_\mathrm{M32}=0.02\,\mathrm{days}$, and 
		and M52 kernels with hyperparameters sampled from the posterior. 
		The similarity metric $\Delta$ is calculated by summing the probability density of the M52 kernel samples along the curve of the true kernel. The settings for the plot are $N_\mathrm{data}=75$ and $\log_{10}(\mathrm{SNR})\approx 0.30$.
		\textbf{Bottom:} For each kernel, $\ln\Delta$ is shown. Larger $\Delta$ corresponds to higher similarity between the inferred and true kernel. It is seen that, for any $t_\mathrm{max}$, the true M32 kernel is not approximated well by the M32 kernel posterior compared to other kernels leading to a flat kernel posterior $p_k$.
	}
	\label{fig:kernel-posterior}
\end{figure}

In addition to the kernel posterior, we investigate how well the reconstructed kernel 
approximates the true kernel. 
For this, we define a similarity metric $\Delta$ which depends on a positive upper cutoff $t_\mathrm{max}$. The calculation thereof is illustrated in Figure~\ref{fig:kernel-posterior}.
Firstly, we draw $N=1000$ hyperparameter posterior samples for a given kernel $k$ and plot the corresponding kernel functions $k(0,t)$ against $t\in [0, t_\mathrm{max}]$.
Next, we discretise the $t$-axis at finely spaced cuts $\{t_i\}$.
At a given cut $t_i$, the plotted kernel functions constitute a probability distribution with density $p_{k,t_i}$, which we compute with a kernel density estimate (KDE; Appendix~\ref{appendix:implementation-of-the-similarity-metric}).
This probability density is evaluated at the value of the M32 kernel, $k_\mathrm{M32}(0,t_i)$.
The value of $\Delta$ is obtained by summing these probability densities over the cuts $\{t_i\}$ and dividing by $N$. 
Therefore, if the probability density of the reconstructed kernel is large along the path traced out by the graph of the true kernel function, corresponding to a higher similarity of between the reconstructed and true kernel, the value of $\Delta$ is large.
Figure~\ref{fig:kernel-posterior} shows $\Delta$ as a function of $t_\mathrm{max}$ for $N_\mathrm{data}=75$ and $\log_{10}(\mathrm{SNR})\approx0.30$.
It is seen that for all $t_\mathrm{max}$, the similarity between the inferred and true M32 kernels is lower than for other kernels. 
Instead, the RQ kernel approximates the M32 kernel more closely.
This is consistent with the results of the kernel posterior, in which the M32 kernel is not the MAP kernel at $N_\mathrm{data}=75$ and ${\log_{10}(\mathrm{SNR})\approx0.30}$, and demonstrates that pre-selecting the M32 kernel does not necessarily reconstruct the ground truth most accurately in comparison to kernel marginalisation.

Finally, we investigate the applicability of these results for the kernel posterior when a mean function is included in the inference.
This not only demonstrates consistency with the method in Section~\ref{sec:kernel-inference-results}, in which the mean function was artificially held fixed, but also shows applicability to real datasets which contain multiple occultations.
For this, the dataset is extended to include up to three occultations. It now consists of multiple copies of the occultation shown in Figure~\ref{fig:mean-function-example} next to each other.
Figure~\ref{fig:multiple-occultations-contour-plots} shows the analogue of Figure~\ref{fig:contour-plot} when the mean function hyperparameters are fit to the data in addition to the kernels.
Note that the number of data points is increased in proportion to the number of occultations contained in the dataset. This keeps the data density on the input domain, i.e. the number of data points per time interval, constant while increasing the size of the domain. This is important as a fit to a free mean function, which increases the dimensionality of the hyperparameter space, requires multiple statistically independent realisations of the same occultation without diluting the data in order to increase the information contained in the dataset.

With a single occultation in the dataset, corresponding to the dataset of this Section, the M32 kernel is correctly inferred at large values of SNR. However, the added complexity by extending the hyperparameter space to include the mean function hyperparameters causes the inference at low SNR to be prior-dominated, verified by inspecting the KL divergence, which specifically results in the preference to the RQ kernel for certain low-SNR datasets. By including a second occultation while fixing the occultation duration, the preference to the RQ kernel disappears, and the separation into low and high entropy kernel posteriors depending on the value of SNR emerges, in agreement with Figure~\ref{fig:contour-plot}. Additionally, the broad kernel posteriors are confined to the lower left triangular region in the plot of $S$ against $N_\mathrm{data}$ and $\log_{10}(\mathrm{SNR})$, which shrinks further as a third occultation is added into the dataset. The red border demarcating that the M32 kernel is the MAP kernel thus extends diagonally from the top left to the bottom right of such a plot. This indicates that increasing the number of data points at a fixed SNR or increasing SNR at a fixed number of data points improves kernel inference if sufficient information is included in the dataset, which in this case means that the domain of inputs is enlarged, which is distinct from keeping the domain of the inputs fixed and increasing the density of data points, which can lead to no constraints on the kernel posterior despite the large $N_\mathrm{data}$ limit, as discussed above. This is also in agreement with the rough notion that more statistically independent data and lower noise generally improve inference. Moreover, the kernel inference remains correct as one moves further into the upper triangular region beyond the red border, indicating that there is a threshold in the number of data points or SNR beyond which there is little improvement in terms of kernel inference, which in practice may serve as a critical number of data points and SNR at which the inference becomes correct.
In summary, we conclude that the results described in this Section are robust if a mean function is included under the condition that the mean function contains more occultations.

We remark that the only remaining difference to Figure~\ref{fig:contour-plot} at low SNR lies in the occasional preference to kernels similar to M32, such as the M52 kernel. This is explained by noting that the mean function inference barely improves as more occultations are included, which was verified by inspection of the hyperparameter posteriors. This is a result of the occultation period, $T=0.12\,\mathrm{days}$, being larger than the kernel length scale, $\ell_\mathrm{M32}=0.02\,\mathrm{days}$, which means that the number of independent samples of the kernel increases more rapidly than the number of independent samples of the mean function as the number of occultations is increased. This implies that the inference of the kernel is overconfident despite wrong mean function inference. A possible way of fixing this would be to increase the number of occultations but simultaneously increase the kernel length scale such that the number of independent kernel samples, approximated by the fraction $\frac{\text{input range}}{\ell_\mathrm{M32}}$, remains constant. 
This approach was not implemented here as it requires one to entirely change the datasets investigated. However, this situation is not unrealistic as noise length scales can be larger than the occultation period in real applications. 
By including a significantly larger number of occultations, while increasing the kernel length scale, it is expected that we approach the limit of a fixed mean function and thus the case investigated in this Section.

In the next Section, mean function hyperparameter inference is investigated on this problem, focussing on a dataset with a single occultation as the hyperparameter inference is relatively unchanged as up to three more occultations are added.

\subsection{Mean function hyperparameter inference results}\label{sec:mean-function-hyperparameter-inference-results}

The agreement of the inferred mean function hyperparameters with their true values is within one sigma. However, the posteriors of the hyperparameters $q_2$ and $b$, which control the precise shape of the occultation, and $P$ remain systematically broad for different $N_\mathrm{data}$ and SNR values due to the lack of constraint from the data as outlined in the previous Section.

\begin{figure*}
		\centering
		\includegraphics{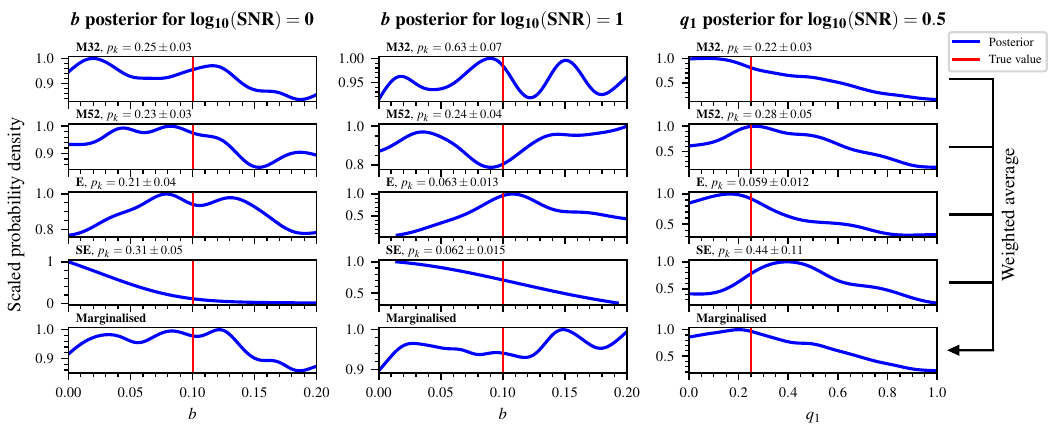}
		\vspace{-2em}
		\caption{Examples of the effect of kernel marginalisation on hyperparameter inference showing that when the true hyperparameters cannot be inferred by both the M32 kernel and the maximum a posteriori kernel, kernel marginalisation still increases the accuracy. Posterior distributions for the exoplanet mean function hyperparameters, $b$ and $q_1$, are shown for synthetic datasets with 
			$N_\mathrm{data}=75$. 
			\textbf{Left:} While the SE kernel, which has the maximum posterior kernel probability $0.31\pm 0.05$ among the kernels, prefers $b=0$, the E kernel is peaked close to the true value despite its smaller posterior kernel probability. In total, the contributions from the M32, M52 and E kernels remove the bias of the SE kernel to low $b$ values, giving a marginalised posterior peaked around the true value. 
			\textbf{Middle:} In the low noise region, the 
			M52 kernel moves posterior mass away from the true value while the marginalised posterior is more uniform, thus representing the lack of information in the data more faithfully.
			\textbf{Right:} Neither the MAP kernel nor the M32 kernel are peaked at the true value. However, the contributions from other kernels give a marginalised posterior peaked close to the true value. 
		}
		\label{fig:mean-function-posteriors}
\end{figure*}

In the cases where the true hyperparameters cannot be precisely inferred due to the broadness of the posteriors, kernel marginalisation shows three favourable effects. These are illustrated in Figure~\ref{fig:mean-function-posteriors} for datasets with $N_\mathrm{data}=75$. Firstly, for the $b$ posterior in the high noise region ($\log_{10}(\mathrm{SNR})=0$), the MAP
kernel is biased to the wrong value, $b=0$. However, contributions from the other kernels, which are significant because the kernel posterior is not sharply peaked, remove this bias, resulting in a marginalised posterior peaked closer to the true value, $b=0.1$. 
Secondly, in the low noise region ($\log_{10}(\mathrm{SNR})=1$), 
inference with the M52 kernel moves posterior mass away from the true $b$ value, whereas the M32 posterior shows little change from the uniform prior as the fluctuations in the posterior density are around $8\%$ of the maximum value. Marginalisation has the effect of smoothing out the M52 posterior and moving posterior mass back to the true $b$ value. Moreover, the E kernel contributes to this as it is peaked at the true $b$ value. The marginalised posterior thus provides a more faithful representation of the lack of information in the data than the M52, E or SE kernels would suggest by themselves.
Finally, for the $q_1$ posterior in the intermediate noise region (${\log_{10}(\mathrm{SNR})=0.5}$), all kernels are peaked at distinct $q_1$ values around the true value. Marginalisation results in a unimodal posterior close to the true value. In contrast to this, inference with the M32 kernel results in a MAP estimate further away from the true $q_1$ value.

\begin{figure}
	\centering
	\includegraphics{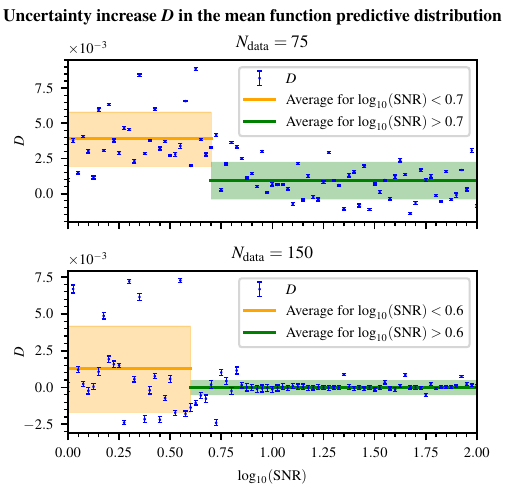}
	\vspace{-3em}
	\caption{Uncertainty increase $D$ in the mean function predictive distribution against the signal-to-noise ratio $\log_{10}(\mathrm{SNR})$ for $N_\mathrm{data}=75$ and $150$. The error bars on each data point measure the 
		variation of this uncertainty over the inputs in a dataset.
		The orange and green lines are the mean of the blue data points in their respective $\log_{10}(\mathrm{SNR})$-ranges and their error bands are the respective one-sigma ranges.
		There is a transition to an average increase in the uncertainty at a critical noise level roughly where the kernel posterior transitions from being flat to peaked at the true kernel (red border in Figure~\ref{fig:contour-plot}).
	}
\label{fig:mean-function-uncertainty-increase}
\end{figure}

Finally, we investigate the effect of kernel marginalisation on the width of the one sigma
error band of the mean function predictive distribution\footnote{The mean function predictive distribution is calculated by drawing samples from the hyperparameter posteriors of the mean function, plotting the corresponding mean functions and calculating the distribution at each input~$t$.}.
For this, we take the difference between the error bands at each input $t$ between the marginalised and M32 predictive distributions and average over all inputs $t$. This defines the uncertainty increase, $D$, which is shown for a series of datasets in Figure~\ref{fig:mean-function-uncertainty-increase}. It is observed that there is a transition from $D\gtrsim 0$ to $D\approx0$ at the value of $\log_{10}(\mathrm{SNR})$ at the red border in Figure~\ref{fig:contour-plot}. This means that the additional uncertainty in the kernel choice is captured when the kernel posterior is flat, whereas the predictive distribution of the true kernel is reproduced when the true kernel maximises the kernel posterior.

In summary, kernel marginalisation performs the inference at least as accurately as using the M32 kernel. This is because in the low noise region, inference results from the true kernel are reproduced whereas in the high noise region, the contributions from other kernels remove bias in the M32 kernel. 
These results should be extended to lower the time complexity, $\varcal{O}(N_\mathrm{data}^3N_\mathrm{dims}^3)$, which warrants improvements when compared to that of ML-II estimates. Therefore, the extension to approximate GP methods \citep{rasmussen} should be investigated.

\section{Hubble parameter inference}\label{sec:cc-data}

In this Section, the method of kernel marginalisation is applied to $H_0$ inference from CC \citep{renzi} and BAO \citep{li} datasets, which consist of measurements of the Hubble parameter $H$ against redshift $z$.

\subsection{Method}\label{sec:hubble-parameter-inference-method}

\begin{figure}
	\centering
	\includegraphics{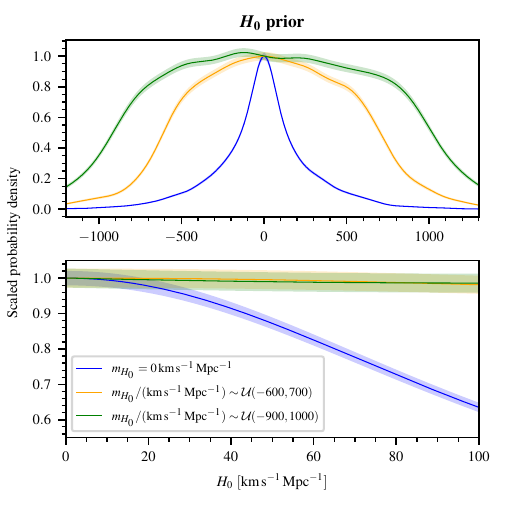}
	\vspace{-3em}
	\caption{Prior probability distributions for $H_0$ induced by the hyperparameter priors for each of the mean function priors in Section~\ref{sec:hubble-parameter-inference-method}.
		The distributions are kernel density estimates with one-sigma
		bootstrap error bands.
		The uniform priors on $m_{H_0}$ broaden the $H_0$ prior and centre it at ${H_0=50\,\mathrm{km}\,\mathrm{s}^{-1}\,\mathrm{Mpc}^{-1}}$, thus removing the bias towards ${H_0=0\,\mathrm{km}\,\mathrm{s}^{-1}\,\mathrm{Mpc}^{-1}}$.
		The shown priors are calculated for the M32 kernel but have the same shape for all other kernels.
}
	\label{fig:H0-priors}
\end{figure}

The $H_0$ posterior is obtained by fitting a GP to an $H(z)$ dataset, extrapolating it to $z=0$ and calculating the marginalised predictive distribution at $z=0$.
All amplitude, $A$, and length scale, $\ell$, hyperparameter priors are set to the conservative ranges $[0\,\mathrm{km}\,\mathrm{s}^{-1}\,\mathrm{Mpc}^{-1},500\,\mathrm{km}\,\mathrm{s}^{-1}\,\mathrm{Mpc}^{-1}]$ and $[0,20]$, respectively. Additionally, the mean function is set to a constant, $m_{H_0}$, for which three priors are investigated:
\begin{enumerate}
	\item [(M.i)] $m_{H_0}=0\,\mathrm{km}\,\mathrm{s}^{-1}\,\mathrm{Mpc}^{-1}$,
	\item [(M.ii)] $m_{H_0}/(\mathrm{km}\,\mathrm{s}^{-1}\,\mathrm{Mpc}^{-1})\sim \varcal{U}(-600,700)$,
	\item [(M.iii)] $m_{H_0}/(\mathrm{km}\,\mathrm{s}^{-1}\,\mathrm{Mpc}^{-1})\sim \varcal{U}(-900,1000)$.
\end{enumerate}
These induce the $H_0$ priors shown in Figure~\ref{fig:H0-priors}. Option (M.i) corresponds to a zero mean function, as done in \citet{gomez, bernardo, yu}, and results in an $H_0$ prior biased towards ${H_0=0\,\mathrm{km}\,\mathrm{s}^{-1}\,\mathrm{Mpc}^{-1}}$. Options (M.ii) and (M.iii) induce an $H_0$ prior which is the convolution of the $H_0$ prior of (M.i) and the prior on $m_{H_0}$, thus broadening the $H_0$ prior and removing the bias. The range of the $m_{H_0}$ priors was chosen such that the induced $H_0$ prior is centred on ${H_0=50\,\mathrm{km}\,\mathrm{s}^{-1}\,\mathrm{Mpc}^{-1}}$ and flat within the range $[0\,\mathrm{km}\,\mathrm{s}^{-1}\,\mathrm{Mpc}^{-1},100\,\mathrm{km}\,\mathrm{s}^{-1}\,\mathrm{Mpc}^{-1}]$, which encompasses all $H_0$ measurements to date \citep{shah}.

Additionally, three contrasting options for the GP noise term were investigated:
\begin{enumerate}
	\item [(N.i)] $\bm{\Sigma}=\mathrm{diag}(\sigma^2_1,\dots,\sigma^2_{N_\mathrm{data}})$, where $\mathrm{diag}$ places its argument along the diagonal of a matrix and $\{\sigma_i\}$ are the measurement error bars provided in the dataset \citep{gomez, bernardo, yu}.
	\item [(N.ii)] $\bm{\Sigma}=\beta^2\mathrm{diag}(\sigma^2_1,\dots,\sigma^2_{N_\mathrm{data}})$, where $\beta$ is a hyperparameter fitted to the dataset with the prior $\beta\sim\varcal{U}(0,5)$. This option assesses whether the sizes of the error bars provided in the dataset are correctly estimated, in which case $\beta=1$ is expected.
	\item [(N.iii)] $\bm{\Sigma}=\sigma^2\mathbf{I}$, where $\sigma$ is a hyperparameter fitted to the dataset with the prior $\sigma/(\mathrm{km}\,\mathrm{s}^{-1}\,\mathrm{Mpc}^{-1})\sim\varcal{U}(0, 500)$.
	This option ignores the measurement error bars in the dataset and fits homoscedastic white noise. It is expected that $\sigma$ is consistent with the size of the error bars $\{\sigma_i\}$ of option (N.i).
\end{enumerate}

In the following, the $H_0$ values for each of the nine choices of mean function priors and noise terms are calculated and, using the evidence $\ln\varcal{Z}$, we marginalise $H_0$ over these modelling choices \citep{scheutwinkel2023bayesian}. This calculation is performed for the CC and BAO datasets and their combination.
Finally, to assess whether the CC and BAO datasets are consistent, a tension analysis \citep{handley2019quantifying} is performed from the entire model consisting of the nine choices and the kernel choice therein.

\subsection{Results}

\begin{figure*}
	\centering
	\includegraphics[width=0.99\textwidth]{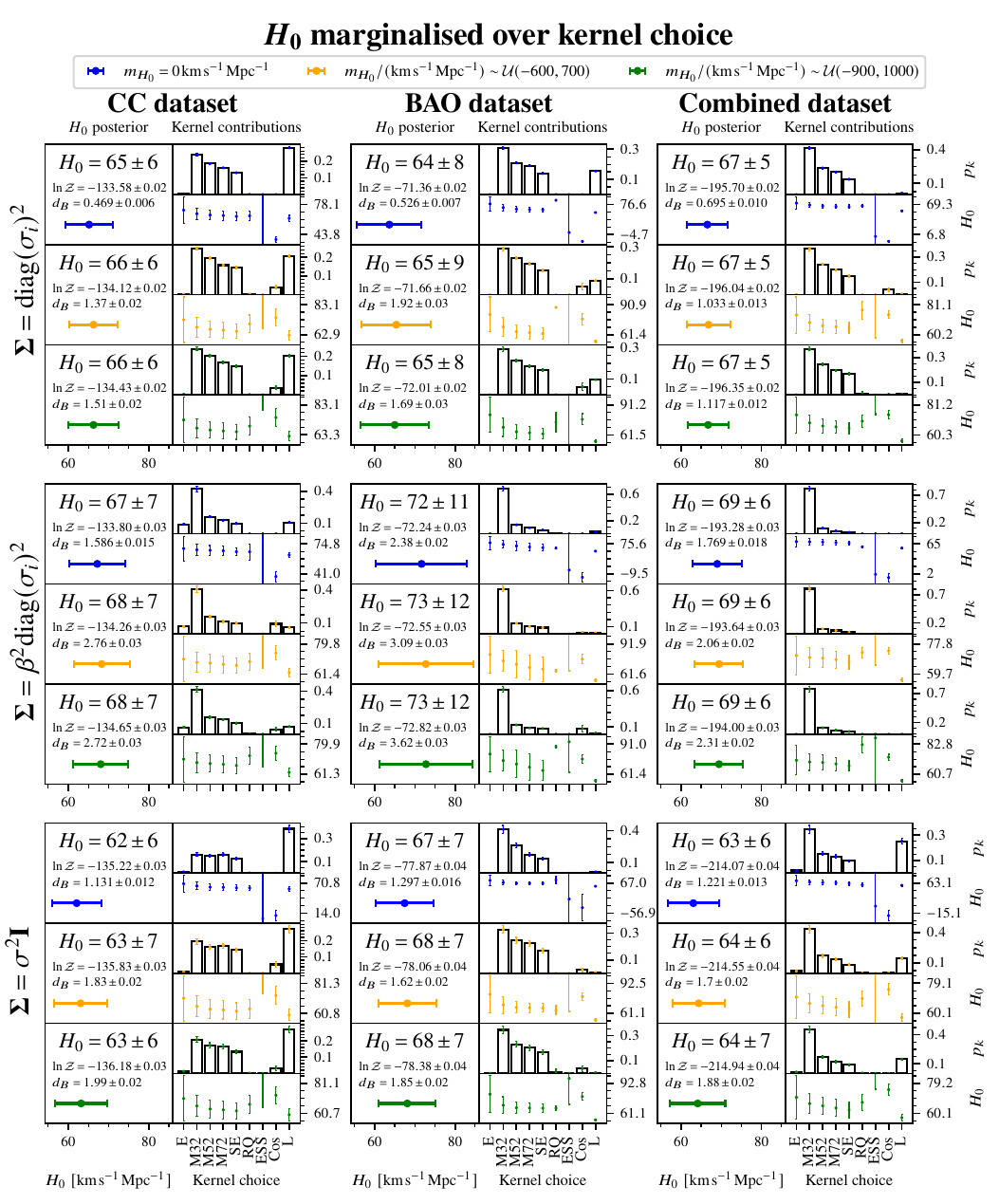}
	\vspace{-1em}
	\caption{
 Means and one-sigma credible intervals of $H_0$, evidences $\ln\varcal{Z}$ and Bayesian model dimensionalities $d_B$ \citep{handley-2} obtained by marginalising over the kernel choice. The columns of the grid of nine subplots correspond to the choice of dataset on which GPR is performed, which are the CC, BAO and the combined CC and BAO datasets. The rows correspond to three different noise models.
 Within each subplot, each colour corresponds to a choice for the $H_0$ prior ($\text{Figure}$~\ref{fig:H0-priors}). The right column shows the kernel probabilities $p_k$ and the $H_0$ values inferred from each kernel. The left column shows the final $H_0$ values obtained by marginalising over these probabilities. It should be noted that due to the width of the error bars of some $H_0$ values of the ESS kernel, their values display as vertical lines on the plot.
  }
	\label{fig:H0-posteriors}
\end{figure*}

The $H_0$ 
means and one-sigma credible intervals
corresponding to the GP fitting options above are shown in Figure~\ref{fig:H0-posteriors}. 
Within each of the nine subplots, the mean $H_0$ 
value marginalised over all kernels is shown. Computationally, this value is produced by picking a kernel according to the kernel posterior, represented by the histogram. Given this kernel, a value of $H_0$ is sampled from the $H_0$ posterior conditioned on the kernel choice shown vertically below the selected kernel. Finally, the mean and one-sigma credible interval is computed from the samples.
Additionally, we report the values of $\ln\varcal{Z}$ and the BMD $d_B$.

The $R$ statistic, information ratio $I$ and suspiciousness $s$, defined in \citet{handley2019quantifying}, evaluate to $\ln R=12.17\pm 0.02$, $\ln I=0.29\pm 0.03$ and ${\ln s=11.88\pm 0.02}$, respectively. Since $R\gg 1$, we conclude that the CC and BAO datasets are consistent for the used prior ranges.

For both datasets, all nine fitting options produce $H_0$ values consistent within one sigma.
Marginalising over these gives the final values ${H_0= 66 \pm 6\, \mathrm{km}\,\mathrm{s}^{-1}\,\mathrm{Mpc}^{-1}}$, ${H_0= 67 \pm 10\, \mathrm{km}\,\mathrm{s}^{-1}\,\mathrm{Mpc}^{-1}}$ and ${H_0= 69 \pm 6 \, \mathrm{km}\,\mathrm{s}^{-1}\,\mathrm{Mpc}^{-1}}$ for the CC, BAO and combined CC and BAO datasets, respectively. 
The evidences of the nine models for each dataset were converted into model posterior probabilities.
For all three datasets, the noise model (N.iii) has negligible posterior probability, contributing at most $9.3\%$ in the CC dataset. This low posterior probability is expected because a homoscedastic model is fitted to heteroscedastic data.
For the CC and BAO datasets, the noise model (N.i) has the largest contribution, accounting for $49.7\%$ and $70.3\%$, respectively, followed by the noise model (N.ii). However, for the combined dataset, noise model (N.ii) contributes dominantly with $91.6\%$.
This preference for (N.ii) causes the shift to larger $H_0$ for the combined dataset, as explained below when (N.i) and (N.ii) are compared. In essence, (N.ii) reduces the size of the measurement error bars which causes the fit to take on a more sigmoid shape in order to follow the data points more closely. This is achieved by a smaller decorrelation length of the kernel which in turn entails a quicker decay to the mean function so that $H_0$, the intercept at $z=0$, becomes larger.

As shown in Figure~\ref{fig:H0-posteriors}, the $H_0$ values conditioned on
the kernels \{E, M32, M52, M72, SE, RQ, ESS, L\} are consistent 
within 
one sigma. For $m_{H_0}=0\,\mathrm{km}\,\mathrm{s}^{-1}\,\mathrm{Mpc}^{-1}$, the inference from the Cos kernel 
shows a larger discrepancy.
However, the Cos kernel contribution to the marginalised $H_0$ value is suppressed by the kernel posterior, $p_k\approx 0$. 
For the two other $H_0$ priors, the Cos kernel agrees with the other kernels within one sigma.
This is because the posterior for the period hyperparameter, $P_\mathrm{Cos}$,
increases up to $P_\mathrm{Cos}=5$ which is larger than the range of redshifts in the dataset so that the effect of the periodicity becomes negligible. For the ESS kernel, the uncertainty $\sigma_{H_0}$ is large, $27<\sigma_{H_0}\,[\mathrm{km}\,\mathrm{s}^{-1}\,\mathrm{Mpc}^{-1}]<125$, compared to the other kernels and the kernel posterior probability, $p_k\approx 0$, suppresses its contribution. These results are consistent with the expectation that periodic kernels are not necessarily suitable to describe non-periodic data. Kernel marginalisation automatically accounts for this. Overall, marginalisation thus removes the inductive bias of the kernel choice.

\begin{figure}
	\centering
	\includegraphics{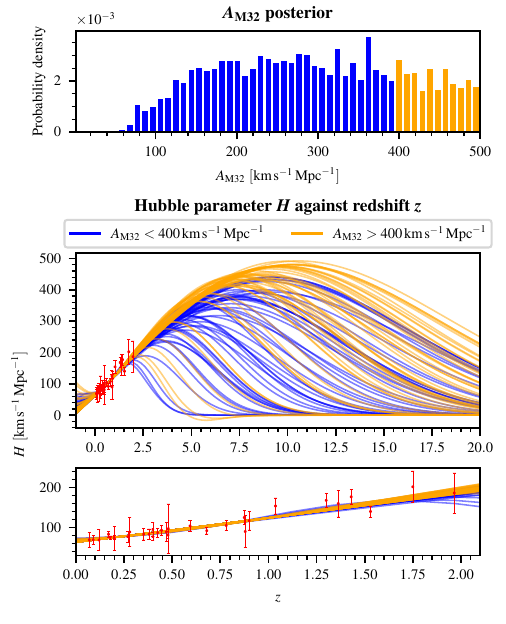}
	\vspace{-3em}
	\caption{\textbf{Top:} Posterior of the amplitude, $A_\mathrm{M32}$, when GPR is performed with the M32 kernel and the mean function set to zero and the noise model $\bm{\Sigma}=\mathrm{diag}(\sigma_i)^2$. The blue and orange regions are separated by $A_\mathrm{M32}=400\,\mathrm{km}\,\mathrm{s}^{-1}\,\mathrm{Mpc}^{-1}$. 
		The tail of the distribution does not agree with the identical posterior calculated in \citet{gomez} despite re-calculation with MCMC methods, which we suspect to be due to the prior choice.
		\textbf{Bottom:} Mean of the predictive distributions corresponding to samples taken from the blue and orange regions of the left subplot. The CC dataset is shown in the 
		lower plot. The orange and blue curves give similar linear fits within the data range but when extrapolated, the orange curves have a wider $H$-range, motivating the use of the linear kernel.}
	\label{fig:posterior-M32-samples}
\end{figure}

For the CC data with noise model (N.iii),
the MAP kernel is consistently the linear kernel. Since functions sampled from a GP with a linear kernel are linear functions, this indicates that linear regression suffices to describe the data. This is validated by plotting the predictive distribution for the M32 kernel separately for ${A_\mathrm{M32}<400\,\mathrm{km}\,\mathrm{s}^{-1}\,\mathrm{Mpc}^{-1}}$ and ${A_\mathrm{M32}>400\,\mathrm{km}\,\mathrm{s}^{-1}\,\mathrm{Mpc}^{-1}}$, showing that the fits through the data points are nearly identical and well approximated by a linear function ($\text{Figure}$ \ref{fig:posterior-M32-samples}). Furthermore, this shows that samples with $A_\mathrm{M32}>400\,\mathrm{km}\,\mathrm{s}^{-1}\,\mathrm{Mpc}^{-1}$ do not change the inferred $H_0$ value and that the chosen prior range is sufficiently wide to infer $H_0$ accurately.
When the BAO measurements are added, $p_k$ of the linear kernel decreases in favour of the M32 kernel. This is because the BAO dataset provides additional $H(z)$ measurements in the ranges $0\lesssim z\lesssim 0.6$ and $2.33<z<2.36$ which are close to the boundary of the CC dataset $z$-range and have smaller error bars, thus preferring a more sigmoid function for which kernels other than the linear kernel are more suited.

Using noise model (N.i)
has four effects when compared to (N.iii).
Firstly, it leads to an increase in $H_0$ (Figure~\ref{fig:H0-posteriors}) for the CC and combined datasets. 
This is because the variation of $H(z)$ near $z=0$ is fitted as uncorrelated noise when (N.iii) is used,
whereas the included error bars in (N.i) constrain the fit to curve upwards around $z=0$, thus lowering the probability of the linear kernel and shifting $H_0$ to higher values.
The same effect causes the addition of the BAO to the CC data to increase $H_0$,
as sigmoid functions become preferred. For the BAO dataset alone, however, changing noise model (N.i) to (N.iii) increases $H_0$ which again correlates with the decrease of posterior probability of the linear kernel.
Secondly, the uncertainty on $H_0$ decreases (increases) for the CC and combined (BAO) datasets, when changing the noise model from (N.iii) to (N.i). In this case, if the posterior probabilities of the linear and M32 kernels are comparable, the error on $H_0$ increases because the linear kernel produces an $H_0$ value smaller than the M32 kernel, albeit with smaller error bars, thus leading to an overall increase in the variance on $H_0$.
Thirdly, the mean of the 
$H_0$ value increases with the width of the $H_0$ prior, i.e. in the order (M.i), (M.ii) and (M.iii), albeit within one sigma. This shows that the prior choice of $m_{H_0}=0\,\mathrm{km}\,\mathrm{s}^{-1}\,\mathrm{Mpc}^{-1}$ contains a bias to low $H_0$ values, as expected from Figure~\ref{fig:H0-priors}. If the constraint on $H_0$ from these datasets were tighter, the bias would be more apparent.
Finally, the increase in prior volume causes the evidence to decrease significantly, making it desirable to find alternative parametrisations of the GP model which are unbiased \citep{handley2019maximum}.

\begin{figure}
	\centering
	\includegraphics{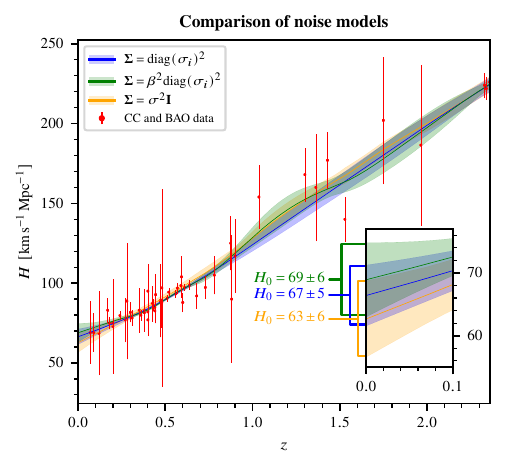}
	\vspace{-3em}
	\caption{
		Kernel-marginalised GP fits for the three options of the noise term, $\bm{\Sigma}$. The option $\bm{\Sigma}=\beta^2\mathrm{diag}(\sigma_i)^2$ effectively fits a datasets with smaller measurement errors. Hence, the GP error band is narrower in the range $z\in[0.1, 0.9]$. However, the error band broadens more strongly than for the other two GP fits in the range $z\in[0, 0.1]$, thus resulting in larger error bars on $H_0$, as shown in the inset. This plot was produced for the combined CC and BAO datasets and the mean function option $m_{H_0}=0\,\mathrm{km}\,\mathrm{s}^{-1}\,\mathrm{Mpc}^{-1}$.
	}
	\label{fig:comparison-of-noise-models}
\end{figure}

Comparing noise models (N.i) and (N.ii), it is observed that (N.ii) has both larger $H_0$ values and a larger uncertainty $\sigma_{H_0}$. Despite the differences in $H_0$ being within the error, this is accounted for by the combination of two effects. Firstly, we note that the noise model (N.ii) fits a value of $\beta$ smaller than $1$. In particular, this is $\beta=0.70\pm0.13$, $\beta=0.72\pm0.16$ and $\beta=0.67\pm0.09$ for the CC, BAO and combined datasets, respectively. Therefore, the scatter in the individual data points are smoothed less by the noise term $\bm{\Sigma}$, which is proportional to $\beta^2$, and must instead be fitted by the kernel. As a result, the decorrelation length scale of the kernel decreases. For example, the M32 kernel length scale $\ell_\mathrm{M32}$ decreases from $10.3\pm 5.2\,\mathrm{km}\,\mathrm{s}^{-1}\,\mathrm{Mpc}^{-1}$ to $7.7\pm 4.4\,\mathrm{km}\,\mathrm{s}^{-1}\,\mathrm{Mpc}^{-1}$ when fitting the combined dataset with mean function (M.i) with noise models (N.i) and (N.ii), respectively. This leads to a strong swelling of the error band of the GP predictive distribution in regions where data is sparse (Figure~\ref{fig:comparison-of-noise-models}), which in turn leads to the larger error bars on the extrapolated $H_0$ values. In addition, the smaller decorrelation length causes the GP to decay to the mean function over a shorter redshift range so that the fit with noise model (N.ii) curves up more strongly in the vicinity of $z=0$, leading to a larger $H_0$ value (Figure~\ref{fig:comparison-of-noise-models}). Secondly, the posterior probability of the linear kernel decreases when the noise model is changed from (N.i) to (N.ii) because $\beta$ being less than $1$ favours fits which follow all data points more closely. We note that the linear kernel produces slightly smaller $H_0$ values compared to the other Mat\'ern kernels (E, M32, M52, M72, SE) because the linear functions extend to $z=\pm\infty$ as a consequence of non-stationarity, whereas the Mat\'ern kernels revert to the mean function. Thus, the disappearance of the preference to the linear kernel also shifts $H_0$ to larger values.

We note that the hyperparameter $\sigma$ in noise model (N.iii) attains an average value of $13\pm2\,\mathrm{km}\,\mathrm{s}^{-1}\,\mathrm{Mpc}^{-1}$ and $9.4\pm1.1\,\mathrm{km}\,\mathrm{s}^{-1}\,\mathrm{Mpc}^{-1}$ in the CC and combined datasets, respectively, which is less than the average error bar size of these datasets, $20\,\mathrm{km}\,\mathrm{s}^{-1}\,\mathrm{Mpc}^{-1}$, and $14\,\mathrm{km}\,\mathrm{s}^{-1}\,\mathrm{Mpc}^{-1}$, suggesting that the measurement errors are overestimated. This conclusion is consistent with the average values of $\beta$ in noise model (N.ii) being less than $1$.
Furthermore, if the white noise term, $\sigma^2\mathbf{I}$, had been included in the covariance matrix of the extrapolation, the error bar size on the inferred $H_0$ values would increase to $14\,\mathrm{km}\,\mathrm{s}^{-1}\,\mathrm{Mpc}^{-1}$ and $11\,\mathrm{km}\,\mathrm{s}^{-1}\,\mathrm{Mpc}^{-1}$ in the two datasets, respectively, which is approximately twice the size of the error bars shown in Figure~\ref{fig:H0-posteriors}. This indicates that a significant amount of the variation in the dataset is absorbed in the white noise term. 

Among the Mat\'ern kernels (E, M32, M52, M72, SE), Figure \ref{fig:H0-posteriors} shows that the posterior probability for the M32 kernel is largest, followed by monotonically decreasing posterior probabilities in the order of the M52, M72 and SE kernels, and lastly the E kernel which has zero posterior probability for most fits. This is investigated by decomposing the kernel evidence, $\ln\varcal{Z}_k$, into the goodness of fit, $\langle\ln\varcal{L}\rangle_{\varcal{P}}$, and the KL
divergence, $\varcal{D}_\mathrm{KL}$, according to the identity $\ln\varcal{Z}_k=\langle\ln\varcal{L}\rangle_{\varcal{P}}-\varcal{D}_\mathrm{KL}$ \citep{hergt}. Firstly, the E kernel has the lowest goodness of fit and largest KL divergence, thus explaining its low posterior probability. Secondly, the minor differences in posterior probabilities between the M32, M52, M72 and SE kernels are accounted for by small differences between the goodness of fits and KL divergences (these differences being approximately between $0.2$ and $2$). For almost all fits\footnote{All fits except for the combination of the BAO dataset, noise model (N.iii) and mean functions (M.ii) and (M.iii), for which the KL divergence decreases.}, the goodness of fit decreases and the KL divergence increases in the order of the kernels listed, thus explaining the trend in posterior probabilities. This increase in the KL divergence may be interpreted as a complexity penalty arising from the increased smoothness of the corresponding GPs\footnote{Equivalently to the kernel formulation, a GP may arise as a solution to a stochastic differential equation \citep{hartikainen, sarkka}, wherein the added complexity lies in the inclusion of higher-order derivatives.}. In this sense, the GP fit with the M32 kernel provides the simplest model and the best fit among the Mat\'ern kernels.

Moreover, the BMDs, which are strictly less than the dimensionality of the fitted models, show that not all hyperparameters are tightly constrained. 
The lack of constraint is also reflected in the uncertainty of the $H_0$ values, which are approximately an order of magnitude larger than those of the Planck ($H_0=67.4\pm 0.5\,\mathrm{km}\,\mathrm{s}^{-1}\,\mathrm{Mpc}^{-1}$; \citet{planck}) and SH0ES ($H_0=73.2\pm 1.3\,\mathrm{km}\,\mathrm{s}^{-1}\,\mathrm{Mpc}^{-1}$; \citet{sh0es}) Collaborations, with which the agreement is within two sigma due to the size of the error bars.
Hence, more CC measurements are needed to constrain $H_0$, in agreement with the conclusions in \citet{zhang}.
Given more measurements, $H_0$ inference using kernel marginalisation should be repeated in the future. 
The decreased uncertainty of $H_0$ will likely result in a stronger dependence on 
the $H_0$ prior choice.
Lastly, we note that the tendency towards lower values in the cosmologically model-independent CC dataset, $H_0\lesssim68\,\mathrm{km}\,\mathrm{s}^{-1}\,\mathrm{Mpc}^{-1}$, is consistent with
\citet{cimatti2023revisiting}.

For further investigation, the datasets can be augmented with additional synthetic data points. This way, the dependence of kernel inference on the number of data points, the range of redshifts covered in the dataset and the noise level can be investigated, similarly to Section~\ref{sec:synthetic-data}, to assess the conditions under which the kernel posterior becomes strongly peaked at a single kernel. This would also enable the definite selection between other non-stationary kernels such as polynomial kernels of higher order.

\section{Conclusions}\label{sec:conclusions}

A novel principled fully Bayesian approach to kernel comparison and kernel marginalisation in Gaussian Process regression is introduced and applied to hyperparameter inference.
The key quantity to calculate for inference from a dataset is the posterior over the kernel choice. This was accomplished by implementing a transdimensional sampler which samples over the kernel posterior and their hyperparameter posteriors in a single run. 
The 
approach is applied to synthetic exoplanet transit data and inference of the Hubble parameter from observational cosmic chronometer plus baryon acoustic oscillations data.

The method infers the true kernel in the low noise region of synthetic data from exoplanet light curve simulations and removes kernel preference in the high noise region. The contributions from multiple kernels increase the accuracy and remove bias in the physical hyperparameter posteriors. Furthermore, single-kernel inference underestimates the uncertainty on the mean function, which increases for kernel marginalisation when transitioning from the low to high noise region.

The method was applied to the inference of the present-day Hubble parameter, $H_0$, from measurements of the Hubble parameter against redshift from cosmic chronometers and baryon acoustic oscillations, yielding 
the values $H_0= 66 \pm 6\, \mathrm{km}\,\mathrm{s}^{-1}\,\mathrm{Mpc}^{-1}$ and ${H_0= 67 \pm 10 \, \mathrm{km}\,\mathrm{s}^{-1}\,\mathrm{Mpc}^{-1}}$,
consistent with the Planck and SH0ES values within two sigma.
Additionally, the contributions from individual kernels agree within one sigma
and kernels whose inferred $H_0$ values deviate more strongly are suppressed by the kernel posterior.

From a machine learning perspective, these investigations 
clearly show the characteristics of this method. Specifically,
the inference of the true kernel strongly correlates with the noise level and the number of data points, captures additional uncertainty in the kernel choice, is capable of reproducing and improving existing analyses, automatically selects the most probable kernel and removes the inductive bias introduced from pre-selecting a kernel, which compellingly improves over the conventional maximum likelihood approach.
Due to these characteristics, strong conclusions can be drawn from the application of the method to astrophysical and cosmological data. 
Mainly, it is insufficient to pre-select a kernel in low data point and high noise datasets such as the cosmic chronometers to infer $H_0$. 
Thus, it is evident that existing analyses which pre-select a kernel need to be verified with kernel marginalisation. 

Future explorations of the method are much encouraged and involve the lowering of the computational runtime by the application of approximative GP methods.
Additionally, the approach can be extended to marginalisation over mean functions and noise terms. From a physical perspective, the aspect of kernel selection facilitates the identification of the most probable model, which in turn guides the physical modelling. Therefore, the method is suitable for any dataset using Gaussian Process regression in which there is only a weak modelling preference.

\section*{Acknowledgements}

The authors thank Alan Heavens for very helpful comments on the manuscript.
N. K. thanks 
the Astrophysics group at the Cavendish Laboratory for laying the foundations of Bayesian inference enabling this project, the provision of the Part III projects of the Natural Sciences Tripos and
Peterhouse for providing computing hardware.
This work was performed using the Cambridge Service for Data Driven Discovery (CSD3), part of which is operated by the University of Cambridge Research Computing on behalf of the STFC DiRAC HPC Facility (www.dirac.ac.uk). The DiRAC component of CSD3 was funded by BEIS capital funding via STFC capital grants ST/P002307/1 and ST/R002452/1 and STFC operations grant ST/R00689X/1. DiRAC is part of the National e-Infrastructure.

\section*{Data Availability}

The CC and BAO datasets used in Section~\ref{sec:cc-data} are available in \citet{renzi} and \citet{li}, respectively.
Code for the transdimensional sampler is available at \url{https://github.com/zwei-beiner/transdimensional_sampler}.

\bibliographystyle{mnras}
\bibliography{example} %

\appendix

\section{Calculation of the posterior probability distribution from the evidence}\label{appendix:posterior-probability-distribution-from-evidence}
Given the evidence values $\{\varcal{Z}_i\}$ for the kernels $i\in\{1,\dots,N\}$, the probability distribution is given by 
\begin{equation}\label{eqn:posterior-probability-in-terms-of-evidence}
	p_i=\frac{\varcal{Z}_i}{\sum_{i=1}^N\varcal{Z}_i},
\end{equation}
for all $i\in\{1,\dots,N\}$, assuming a uniform prior over the kernel choices.

Since $\ln\varcal{Z}_i$ is approximately normally distributed \citep{ashton}, the standard deviation, $\sigma_{p_i}$, of $p_i$ in terms of the standard deviation $\sigma_{\ln\varcal{Z}_i}$ can be calculated using Gaussian error propagation and is given by
\begin{align}
	\sigma_{p_i}&=\sqrt{\sum_{j=1}^N\left(\frac{\partial p_i}{\partial(\ln\varcal{Z}_j)}\sigma_{\ln\varcal{Z}_j}\right)^2}\\
	&=p_i\sqrt{(\sigma_{\ln\varcal{Z}_i})^2+p_i^2\sum_{\substack{j=1\\j\neq i}}^N\left(\frac{\sigma_{\ln\varcal{Z}_j}\varcal{Z}_j}{\varcal{Z}_i}\right)^2}\label{eqn:error-in-posterior-probability-in-terms-of-evidence}.
\end{align}

Equations \ref{eqn:posterior-probability-in-terms-of-evidence} and \ref{eqn:error-in-posterior-probability-in-terms-of-evidence} were used if $\{\varcal{Z}_i\}$ and $\{\sigma_{\ln\varcal{Z}_i}\}$ had been pre-computed. Alternatively, these can be computed by sampling from $\ln\varcal{Z}_i$ using the \textsc{anesthetic} library \citep{handley-anesthetic}.

\section{Mean and covariance of a multivariate Gaussian mixture distribution}\label{appendix:multivariate-gaussian-mixture-distribution}

The calculation of any quantity, $Q$, dependent on a kernel and its hyperparameters, such as the Hubble parameter inferred from the CC dataset, was performed using hierarchical sampling from $\text{Equation}$~\ref{eqn:inference-of-quantity}. That is, as described in $\text{Equation}$ \ref{eqn:quantity-sample-from-distribution}, one samples from the joint posterior and then samples from the distribution $p(Q\mid k^{(j)},\bm{\Theta}^{(j)})$ given the sample $(k^{(j)},\bm{\Theta}^{(j)})$. Alternatively, if the distributions $p(Q\mid k^{(j)},\bm{\Theta}^{(j)})$ are Gaussian, such as for the predictive distribution ($\text{Equation}$ \ref{eqn:predictive-distribution}), expressions for the mean and covariance can be derived analytically, which in practice allows fewer samples to be taken.

Let $\bm{x}$ be a random vector with probability density 
\begin{equation}
	p(\bm{x})=\sum_{i=1}^N\pi_i\mathcal{N}_i(\bm{x}),
\end{equation}
where $\mathcal{N}_i=\mathcal{N}(\bm{x}|\mu_i,\Sigma_i)$ are normal distributions with mean vector $\mu_i$ and covariance matrix $\Sigma_i$. Furthermore, $\sum_{i=1}^N\pi_i=1$ and ${0\le\pi_i\le1}$ for all $i$. 

Let $H(\bm{x})$ be an arbitrary function of $\bm{x}$. Then the expectation value of $H(\bm{x})$ is
\begin{align}
	\mathbb{E}_{\bm{x}\sim p}[H(\bm{x})]&=\int H(\bm{x})\sum_{i}\pi_i\mathcal{N}_i(\bm{x})\mathrm{d}\bm{x}\\
	&=\sum_i\pi_i\int H(\bm{x})\mathcal{N}_i(\bm{x})\mathrm{d}\bm{x}\\
	&=\sum_i\pi_i\mathbb{E}_{\bm{x}\sim\mathcal{N}_i}[H(\bm{x})],
\end{align}
where $\mathbb{E}_{\bm{x}\sim p}$ denotes the expectation value with respect to $\bm{x}$, which is sampled from the probability distribution $p$. Hence, the mean of $\bm{x}$ is 
\begin{align}
	\mathbb{E}_{\bm{x}\sim p}[\bm{x}]&=\sum_{i}\pi_i\mathbb{E}_{\bm{x}\sim\mathcal{N}_i}[\bm{x}]\\
	&=\sum_i\pi_i\mu_i.
\end{align}
Using $\mathbb{E}_{\bm{x}\sim\mathcal{N}_i}[\bm{x}\bm{x}^T]=\Sigma_i+\mu_i\mu_i^T$,
\begin{align}
	\mathbb{E}_{\bm{x}\sim p}[\bm{x}\bm{x}^T]&=\sum_{i}\pi_i\mathbb{E}_{\bm{x}\sim\mathcal{N}_i}[\bm{x}\bm{x}^T]\\
	&=\sum_i\pi_i(\Sigma_i+\mu_i\mu_i^T).
\end{align}
Thus, the covariance matrix of $\bm{x}$ is 
\begin{align}
	\mathrm{Cov}_{\bm{x}\sim p}[\bm{x},\bm{x}]
	&=\mathbb{E}_{\bm{x}\sim p}[\bm{x}\bm{x}^T]-\mathbb{E}_{\bm{x}\sim p}[\bm{x}]\mathbb{E}_{\bm{x}\sim p}[\bm{x}]^T\\
	&=\sum_i\pi_i(\Sigma_i+\mu_i\mu_i^T)-\left(\sum_i\pi_i\mu_i\right)\left(\sum_j\pi_j\mu_j\right)^T\\
	&=\sum_i\pi_i\left(\Sigma_i+\mu_i\mu_i^T-\sum_j\pi_j\mu_i\mu_j^T\right).
\end{align}
In the case of the predictive distributions (Equation \ref{eqn:predictive-distribution}), if one has $M$ samples from the joint posterior over kernels and hyperparameters and computes the means, $\mu(\mathbf{x}_i^\star)$, variances, $\sigma(\mathbf{x}_i^\star)^2$, at each training input, $\mathbf{x}_i^\star$, then these expressions reduce to 
\begin{align}
	\text{Mean}&=\langle \mu(\mathbf{x}_i^\star)\rangle,\\
	\text{Standard deviation}&=\left[\langle \sigma(\mathbf{x}_i^\star)^2\rangle+\text{Var}[\mu(\mathbf{x}_i^\star)]\right]^{1/2},
\end{align}
where $\langle\cdot\rangle=\frac{1}{M}\sum_{i=1}^M(\cdot)$ and $\text{Var}[\cdot]=\langle\cdot^2\rangle-\langle\cdot\rangle^2$.

\section{Equivalence of a linear kernel and a linear mean function}\label{appendix:equivalence-of-linear-kernel-and-linear-mean}
In the following, we show that GP regression on one-dimensional inputs $x\in\mathbb{R}$ with a zero mean function and a linear kernel, ${k(x,x')=A_{\mathrm{L},1}^2+A_{\mathrm{L},2}^2xx'}$, is equivalent to linear regression, $f(x)=mx+b$, with a Gaussian prior, $(m\ \ b)^{\top}\sim\mathcal{N}(\mathbf{0},\mathrm{diag}(A_{\mathrm{L},2}^2,A_\mathrm{\mathrm{L},1}^2))$.
As a corollary, functions sampled from the GP are straight lines.

\citet{rasmussen} show that a Bayesian linear regression model $f(x)=\bm{\phi}(x)^{\top}\mathbf{w}$ with a set of basis functions $\bm{\phi}$ and prior $\mathbf{w}\sim\mathcal{N}( \mathbf{0},\bm{\Sigma}_p)$ is a Gaussian process with a zero mean function and covariance $\mathbb{E}[f(x)f(x')]=\bm{\phi}(x)^{\top}\bm{\Sigma}_p\bm{\phi}(x')$. By substituting $\bm{\phi}(x)=(x\ \ 1)^{\top}$, $\mathbf{w}=(m\ \ b)^{\top}$ and $\bm{\Sigma}_p=\mathrm{diag}(A_{\mathrm{L},2}^2,A_{\mathrm{L},1}^2)$, we obtain the linear kernel with $k(x,x')=\mathbb{E}[f(x)f(x')]$, as required.

Alternatively, the equivalence may be proved by directly showing the equality of the distributions of model outputs $\mathbf{f}$, which is a column vector of $f(x)$ applied to $n$ inputs. First, observe that this may be written as $\mathbf{f}=\mathbf{X}(m\ \ b)^{\top}$, where $\mathbf{X}\in \mathbb{R}^{n\times 2}$ is a matrix with the first column filled with the inputs and the second column consisting of ones. 
By the affine transformation property \citep{bierens2004introduction}, it follows that ${\mathbf{f}\sim \mathcal{N}(\mathbf{0},\mathbf{X}\mathrm{diag}(A_{\mathrm{L},2}^2,A_{\mathrm{L},1}^2)\mathbf{X}^{\top})}$. 
By direct evaluation of the matrix product, the covariance matrix of $\mathbf{f}$ can be shown to have entries $k(x_i,x_j)$, as required. 

\section{Example of a uniform categorical prior}\label{appendix:example-of-a-uniform-categorical-prior}

In the following, it is explicitly shown that the categorical uniform prior specified by ${u\mapsto \lceil N_\mathrm{kernel} u\rceil}$ gives the correct evidence. Without loss of generality, suppose that we aim to sample from two kernels and there are no other hyperparameters, i.e. $\mathbf{\Phi}=(c)$ and $c\in \{1, 2\}$. We set a uniform prior so that $\pi(c)=\frac12$ for all $c$ and the likelihood takes on values 
\begin{equation}
	\varcal{L}(c)=
	\begin{cases}
		\varcal{L}_1&\text{if }c=1\\
		\varcal{L}_2&\text{if }c=2
	\end{cases}.
\end{equation}

We now define the inverse transform on the unit hypercube as $F^{-1}(u)=\lceil 2u\rceil$.
In the unit hypercube, the evidence is calculated as 
\begin{align}
	\varcal{Z}&=\int_0^1 \varcal{L}(F^{-1}(u))\mathrm{d}u\\
	&=\int_0^{1/2}\varcal{L}(F^{-1}(u))\mathrm{d}u+\int_{1/2}^1 \varcal{L}(F^{-1}(u))\mathrm{d}u\\
	&=\frac12 \varcal{L}_1+\frac12\varcal{L}_2=\sum_{c=1}^2\varcal{L}(c)\pi(c),
\end{align}
as required.

\section{Testing the transdimensional sampler}\label{appendix:testing-the-transdimensional-sampler}

Tests for the TS were performed with exoplanet transit light curve data and cosmic chronometers (CC) data, described in Sections \ref{sec:synthetic-data} and \ref{sec:cc-data}, respectively. 

\subsection{Evidence calculation}\label{sec:evidence-calculation}

The evidence calculation of the TS was tested on an exoplanet dataset (Section~\ref{sec:synthetic-data}) with the hyperparameters set as in $\text{Table}$~\ref{table:exoplanet-simulation-parameters}, white noise $\sigma=0.001$ and the number of data points $N_\mathrm{data}=75$. 
Two methods were compared:
\begin{enumerate}
	\item ``Two-step'': The evidences $\varcal{Z}_\text{M32}$, $\varcal{Z}_\text{SE}$ and $\varcal{Z}_\text{ESS}$ were calculated directly for the model corresponding to each kernel by calculating the posterior over the hyperparameters and over the kernels in two steps. The evidence $\varcal{Z}$ was obtained using $\text{Equation}$ \ref{eqn:kernel-evidence}.
	\item ``TS'': Using the TS, 
	$\varcal{Z}_\text{M32}$, $\varcal{Z}_\text{SE}$, $\varcal{Z}_\text{ESS}$ and $\varcal{Z}$ were calculated.
\end{enumerate}
The inference with method (i) was performed with $500$ live points. As the live points are partitioned between the kernels, the inference with method (ii) was performed with $4000$ live points. For both methods, a GP model with zero mean function was used and the kernel amplitude was set to $A=1$. The prior $\mathrm{Loguniform}(e^{-20},e^{20})$ was used for the remaining kernel hyperparameters $\ell_\mathrm{M32}$, $\ell_\mathrm{SE}$, $\ell_\mathrm{ESS}$, $\Gamma$ and the white noise $\sigma$. The resulting evidences agreed within the error ($\text{Table}$ \ref{table:testing-evidence-calculation-results}).

\begin{table}
	\centering
	\caption{Results of the evidence calculations using synthetic data from exoplanet transit light curve simulations.
		The difference between the evidences from the two-step and TS methods agree within one sigma.
	}
	\label{table:testing-evidence-calculation-results}
	\begin{tabular}{cccc}
		\hline 
		\textbf{Evidence}&\textbf{Two-step}&\textbf{TS}&\textbf{Difference}\\
		\hline 
		$\ln\varcal{Z}_\text{M32}$&$342.47\pm0.13$&$342.43\pm0.07$&$\mathbf{0.4}\bm{\sigma}$\\
		$\ln\varcal{Z}_\text{SE}$&$316.85\pm0.13$&$316.89\pm0.13$&$\mathbf{0.15}\bm{\sigma}$\\
		$\ln\varcal{Z}_\text{ESS}$&$331.37\pm0.15$&$331.42\pm0.24$&$\mathbf{0.2}\bm{\sigma}$\\
		$\ln\varcal{Z}$&$341.4\pm0.2$&$341.25\pm0.05$&$\mathbf{0.03}\bm{\sigma}$\\
		\hline 
	\end{tabular}
\end{table}

\subsection{Checking the independence of $\varcal{Z}$ on unconstrained hyperparameters}

The TS assumes that for a given kernel choice $c$, 
all unused hyperparameters in $\bm{\Phi}$
retain a uniform distribution \citep{hee}. As the evidence $\varcal{Z}_k$ and Bayesian model dimensionality \citep{handley-2} are unaffected if parameters with a uniform distribution are added into a model, they should be equal if they are calculated from $\bm{\Phi}$ or by slicing out the used hyperparameters
from $\bm{\Phi}$. On the dataset of Section \ref{sec:evidence-calculation}, these gave identical results.

\subsection{Comparison between a linear kernel and a linear mean function}

Using the CC dataset, $H_0$ was inferred using the following GP models:
\begin{enumerate}
	\item $m(x)=0$ and a linear kernel,
	\item a linear mean function and the kernel set to zero,
\end{enumerate}
setting the noise term in both models to the measurement errors from the dataset. The results give $1.5\sigma$ agreement ($\text{Table}$~\ref{table:testing-comparison-between-linear}), which can be decreased further by increasing the number of live points in the nested sampling run.

 \begin{table}
	 	\centering
	 	\caption{Results of the evidence calculations using the cosmic chronometers dataset. 
 		}
	 	\begin{tabular}{ccc}
		 		\hline 
		 		Method&$\ln\varcal{Z}$&$H_0\ [\mathrm{km}\,\mathrm{s}^{-1}\,\mathrm{Mpc}^{-1}]$\\
		 		\hline
		 		Linear kernel&$-132.8\pm 0.3$&$62.1\pm 7.2$\\
		 		Linear mean function&$-132.36\pm 0.12$&$61.3\pm 7.1$\\
		 		\textbf{Difference}&$\mathbf{1.5}\bm{\sigma}$&$\mathbf{0.2}\bm{\sigma}$\\
		 		\hline 
		 	\end{tabular}
	 	\label{table:testing-comparison-between-linear}
	 \end{table}

\section{Prior ranges for kernel hyperparameters}\label{appendix:prior-ranges}

The ranges for the uniform priors used for the kernel hyperparameters 
are shown in $\text{Table}$ \ref{table:prior-ranges}.
In the following, the range for the hyperparameter $\Gamma$ is derived.

\begin{table*}
	\centering
	\caption{Ranges of the uniform priors used for the kernel hyperparameters of the kernels in $\text{Table}$ \ref{table:kernels}. $y_i$, $\mathbf{x}_i$ and $N_\mathrm{data}$ are defined in Section \ref{sec:gaussian-process-regression}. It is assumed that $\mathbf{x}_i$ is a real number.}
	\label{table:prior-ranges}
	\begin{tabular}{cp{2cm}cp{8cm}}
		\hline
		\textbf{Hyperparameter}&\multicolumn{1}{c}{\textbf{Kernels}}&\textbf{Range}&\multicolumn{1}{c}{\textbf{Explanation}}\\
		\hline 
		Amplitude, $A$&M32, M52, M72, RQ, E, SE, ESS, Cos&$\left[0, \sqrt{\frac{\max y_i-\min y_i}{2}}\right]$&$A$ is the square root of the deviation from the mean function. Hence, the upper limit is determined by the square root of half the $y$-range.\\
		Length scale, $\ell$&M32, M52, M72, RQ, E, SE, ESS, Cos&$\left[0, \max \mathbf{x}_i-\min \mathbf{x}_i\right]$&$\ell$ is the correlation length scale of the inputs. Hence, the upper limit is determined by the maximum correlation length in the data, the $\mathbf{x}$-range.\\
		$\alpha$&RQ&$\left[0, 10^{15}\right]$&To allow for the possibility that the RQ kernel converges to the SE kernel for $\alpha\rightarrow\infty$ \citep{rasmussen}, the upper limit is set to the largest number such that the prior probability density, $\left(\text{upper limit}-\text{lower limit}\right)^{-1}$, is larger by one order of magnitude than 
		the machine epsilon 
		for \textsc{numpy}.\\
		$P$&ESS, Cos&$\left[\mathbf{x}_1-\mathbf{x}_0, \frac{\max\mathbf{x}_i-\min \mathbf{x}_i}{2}\right]$&The minimum value of $P$ is determined by the distance between two inputs, $\mathbf{x}_1-\mathbf{x}_0$ \citep{simpson}. The maximum value is given by half of the range of inputs.\\
		$\Gamma$&ESS&$\left[\frac{1}{2\pi^2N_\mathrm{data}^2}, 10^{15}\right]$&See $\text{Appendix}$ \ref{appendix:prior-ranges}.\\
		\hline 
	\end{tabular}
\end{table*}

For large values of $P_\text{ESS}$, the ESS kernel tends to the SE kernel,
\begin{align}
	k_\text{ESS}&=A_\text{ESS}^2\exp\left(-\Gamma\sin^2\left(\frac{\pi\delta}{P_\text{ESS}}\right)\right)\\
	&\approx A_\text{ESS}^2\exp\left(-\Gamma\frac{\pi^2\delta^2}{P^2_\text{ESS}}\right),
\end{align}
with length scale $\ell_\text{SE}=\frac{P_\text{ESS}^2}{2\Gamma\pi^2}$. When regression on non-periodic data with the ESS kernel is performed, it is expected that the posterior of $P_\text{Cos}$ is peaked at large values. Requiring consistency with the SE kernel, the range of $\Gamma$ is thus given by the attainable values of 
\begin{equation}
	\Gamma=\frac{P^2_\text{ESS}}{2\ell_\text{SE}\pi^2}.
\end{equation}
With the ranges of $P_\text{ESS}$ and $\ell_\text{SE}$ from $\text{Table}$ \ref{table:prior-ranges}, this gives the range $\left[\frac{1}{2\pi^2N_\mathrm{data}^2}, \infty\right]$, where $n$ is the number of data points. By the same reasoning as for the hyperparameter $\alpha$ ($\text{Table}$ \ref{table:prior-ranges}), this becomes $\left[\frac{1}{2\pi^2N_\mathrm{data}^2}, 10^{15}\right]$.

\section{Implementation of the similarity metric}\label{appendix:implementation-of-the-similarity-metric}

The calculation of $\Delta(t_\mathrm{max})$ in Section~\ref{sec:kernel-inference-results} requires the calculation of the probability density, $p_{k,t_i}$. This was calculated as a kernel density estimate (KDE) using \textsc{scipy} \citep{scipy}. 
However, the numerical problem arises that as $t$ becomes large, the kernel $k$ tends to zero so that the set of samples contains duplicate values and a KDE cannot be computed. To mitigate this problem, $p_{k,t_i}$, is modelled as the sum of Dirac-delta functions at the $N_d$ duplicate values, ${\{t^{(d)}_i\mid i=1,\dots,N_d\}}$, with multiplicities $\{N_{d_i}\mid i=1,\dots,N_d\}$, and a probability density, $p_\mathrm{KDE}$, computable with a KDE due to the remaining $N_u$ unique values:
\begin{equation}\label{eqn:marginalised-kernel-posterior-kde}
	p_{k,t_i}=\frac{N_u}{N}p_\mathrm{KDE}+\sum_{i=1}^{N_d}\frac{N_{d_i}}{N}\delta(t-t_i).
\end{equation}

\bsp	%
\label{lastpage}
\end{document}